\documentclass[12pt,bezier,amsfonts]{article}
\usepackage{amssymb}
\normalbaselines

\catcode `\@=11
\@addtoreset{equation}{section}

\def\section{\@startsection {section}{1}{\z@}{-3.5ex plus -1ex minus 
 -.2ex}{2.3ex plus .2ex}{\large\bf}}
\def\subsection{\@startsection{subsection}{2}{\z@}{-3.25ex plus -1ex minus 
 -.2ex}{1.5ex plus .2ex}{\normalsize\bf}}
\def\subsubsection{\@startsection{subsubsection}{3}{\z@}{-3.25ex plus
 -1ex minus -.2ex}{1.5ex plus .2ex}{\normalsize\sl}}
\catcode `\@=12

\font\Eul = eufm7 at 12pt 
\font\eul = eufm7 at 6pt

\newcommand{\be}{\begin{equation}}
\newcommand{\en}{\end{equation}}
\newcommand{\bea}{\begin{eqnarray}}
\newcommand{\ena}{\end{eqnarray}}
\newcommand{\beano}{\begin{eqnarray*}}
\newcommand{\enano}{\end{eqnarray*}}
\newcommand{\bee}{\begin{enumerate}}
\newcommand{\ene}{\end{enumerate}}
\newcommand{\bei}{\begin{itemize}}
\newcommand{\eni}{\end{itemize}}

\newtheorem{theorem}{Theorem}[section]
\newtheorem{coroll}[theorem]{Corollary}
\newtheorem{lemma}[theorem]{Lemma} 
\newtheorem{prop}[theorem]{Proposition}  
\newtheorem{defin}[theorem]{Definition}  

\newcommand{\betheo}{\begin{theorem}}
\newcommand{\entheo}{\end{theorem}}

\newcommand{\becor}{\begin{coroll}}
\newcommand{\encor}{\end{coroll}}

\newcommand{\belem}{\begin{lemma}}
\newcommand{\enlem}{\end{lemma}}

\newcommand{\beprop}{\begin{prop}}
\newcommand{\enprop}{\end{prop}}

\newcommand{\beee}{\begin{defin}}
\newcommand{\enee}{\end{defin}}

\newcommand{\proof}{{\em Proof. }-- }
\newcommand{\enproof}{\hfill {$\square$} \\}

\def\BC{{\mathbb C}}
\def\BR{{\mathbb R}}
\def\BN{{\mathbb N}}
\def\BZ{{\mathbb Z}}

\def\B{\relax\ifmmode {\cal B}\else${\cal B}$\fi}
\def\C{\relax\ifmmode {\cal C}\else${\cal C}$\fi}
\def\D{\relax\ifmmode {\cal D}\else${\cal D}$\fi}
\def\F{\relax\ifmmode {\cal F}\else${\cal F}$\fi}
\def\H{\relax\ifmmode {\cal H}\else${\cal H}$\fi}
\def\K{\relax\ifmmode {\cal K}\else${\cal K}$\fi}
\def\L{\relax\ifmmode {\cal L}\else${\cal L}$\fi}
\def\I{\relax\ifmmode {\cal I}\else${\cal I}$\fi}
\def\J{\relax\ifmmode {\cal J}\else${\cal J}$\fi}
\def\O{\relax\ifmmode {\cal O}\else${\cal O}$\fi}
\def\S{\relax\ifmmode {\cal S}\else${\cal S}$\fi}

\def\x{\relax\ifmmode {\mbox{*}}\else*\fi}

\newcommand{\A}{\mbox{\Eul A}}
\newcommand{\M}{\mbox{\Eul M}}
\newcommand{\N}{\mbox{\Eul N}}
\newcommand{\ga}{\mbox{\eul A}}
\newcommand{\n}{\mbox{\eul N}}
\newcommand{\m}{\mbox{\eul M}}

\newcommand{\gK}{\mbox{\Eul K}}

\newcommand{\gP}{\mbox{\Eul P}}

\newcommand{\ha}{^{\rm\textstyle *}}
\newcommand{\haa}{^{\rm\textstyle **}}
\newcommand{\ad}{^{\mbox{\scriptsize{\dag}}}}

\newcommand{\w}{_{\rm w}}

\newcommand{\la}{\lambda}

\newcommand{\LDH}{{\L}\w\ad(\D,\H)}
\newcommand{\LD}[2]{{\L}\ad(#1,#2)}
\newcommand{\LL}[1]{{\cal L}\ad(#1)}
\newcommand{\mult}{\,{\scriptstyle \square}}

\newcommand{\xa}{\mbox{*-algebra}}
\newcommand{\xas}{\mbox{*-algebras}}
\newcommand{\qxa}{\mbox{quasi *-algebra}}
\newcommand{\pa}{partial \mbox{*-algebra}}
\newcommand{\po}{partial O\mbox{*-algebra}}
\newcommand{\pg}{partial GW\mbox{*-algebra}}
\newcommand{\tpa}{topological partial \mbox{*-algebra}}
\newcommand{\vna}{von Neumann algebra}

\newcommand{\up}{\raisebox{0.7mm}{$\upharpoonright$}}

\begin{document}

\thispagestyle{empty}

\begin{center}
{\large\bf Topological partial *-algebras: Basic properties and examples} \vspace{5mm} 
\\
\vspace{2cm}

{\large J.-P. Antoine$^{1}$,
F. Bagarello$^{2}$ and C. Trapani$^{3}$ } \vspace{3mm} \\
$^{1}$ Institut de Physique Th\'eorique, Universit\'e Catholique de Louvain\\
B-1348  Louvain-la-Neuve, Belgium \vspace{2mm}\\ 

$^{2}$ Dipartimento di Matematica dell' Universit\`{a} di Palermo\\ 
I-90123 Palermo, Italy

$^{3}$ Istituto di Fisica dell' Universit\`{a} di Palermo\\ 
I-90123 Palermo, Italy
\end{center}

\vspace{1cm}

\begin{abstract}

Let \A\ be a partial *-algebra endowed with a topology $\tau$ that makes it
into a locally convex topological vector space $\A[\tau]$. Then \A\ is called a 
topological partial *-algebra  if it satisfies a number of conditions, which all
amount to require that the topology $\tau$ fits with the multiplier 
structure of \A. Besides the obvious cases of  topological quasi *-algebras
and CQ*-algebras, we examine several classes of potential
topological partial *-algebras, either function spaces (lattices of
$L^p$ spaces on $[0,1]$ or on $\BR$, amalgam spaces), or 
partial *-algebras of operators (operators on a partial inner product space, 
O*-algebras).

\vfill

\bigskip
\noindent
E-mail: Antoine@fyma.ucl.ac.be \\ 
\hspace*{13mm} 
Bagarello@ipamat.math.unipa.it 
\\ 
\hspace*{13mm} 
Trapani@mbox.unipa.it 
\bigskip\bigskip

\begin{flushright}
 UCL-IPT-97-16\\ November 1997 
\end{flushright}

\end{abstract}

\newpage


\section{\hspace{-5mm}. Introduction and motivation}

A \pa\ is a vector space equipped with a multiplication that is only defined for 
certain pairs of elements. Many different species have cropped up in the recent
mathematical literature, for instance, \qxa s \cite{lass3,lass4}, CQ\xas\ 
\cite{bagtrap1,bagtrap2} or various kinds of \pa s of operators in Hilbert spaces, the
so-called \po s \cite{ak}-\cite{aitrev}.
In all cases, there is an algebraic backbone, the abstract \pa, mentioned in 
\cite{bor3}  and developed in \cite{ak} and \cite{ait1}. 
On top of that, a number of topological properties are introduced. For instance,
\po s were envisaged as generalizations of \xas\ of bounded operators (\vna s or C\xas)
and of \xas\ of unbounded operators or O\xas\ \cite{schm}.

Yet one element is missing in this picture, an abstract notion of \tpa, that
would encompass and unify all these examples. That such a concept is useful is
illustrated by the following situation.

Let $(\A, \A_o)$ be a noncomplete topological \qxa, that is, $\A_o$ is a topological
\xa, but the multiplication is only separately, not jointly continuous for the
topology of $\A_o$, and the latter is not complete. If \A\ is the completion of $\A_o$,
then it is  no longer an algebra in general, but only a partial algebra: a product
$AB$ is defined only (by continuity) if either $A$ or $B$ belongs to $\A_o$.
Let now $\pi_o$ be a *-representation of $\A_o$ by operators acting on a dense domain
$\D(\pi_o)$ in a Hilbert space \H.  This means, in particular, that $\pi_o$ maps $\A_o$
into $\LL{\D(\pi_o)}$, i.e. the space of all closable operators $A$
in \H\ with domain  $D(A) = \D(\pi_o)$ and leaving it invariant. 
Now it is legitimate to ask how one could  extend the representation $\pi_o$
from $\A_o$ to \A, or at least to some larger subset of it.
The obvious way would be by taking limits, using some notion of {\em closability} of 
the representation $\pi_o$. But this implies in general extending $\pi_o$ beyond 
$\D(\pi_o)$, since the extended operators need no longer map $\D(\pi_o)$ into itself. In  \cite{ant-bag-trap} we have perfromed such an extension, by operators in $\L(\D(\pi_o),\D')$, where $\D'$ is the dual of $\D(\pi_o)$
in a suitable topology. However, from the point of view of \po s, a more natural framework for the extension is the space 
$\LD{\D(\pi_o)}{\H}$  of all closable operators $A$
in \H\ such that $D(A) = \D$ and $D(A\ha) \supset \D$, which is a \pa. 
However, in order to perform such an extension by closure, one 
clearly needs a more sophisticated topological
structure on $\LD{\D(\pi_o)}{\H}$ than the one available in the current literature.
As a matter of fact, a large number of interesting results have been obtained
concerning  \po s, such as structure results, (GNS) representations, automorphisms and
derivations (see \cite{aitrev} for a review and references to the original papers),
but the interplay between the (partial) algebraic structure and the topological
properties of \po s has been largely ignored.

It is the aim of the present paper to try and fill this gap. In other words, we want
to find a working definition of {\em \tpa} that would cover all the cases mentioned at
the beginning. Actually these fall into three categories.
\bei

\item[(i)]

Simple cases, such as \qxa s and CQ\xas, whose structure of \pa\ is almost trivial ---
but, of course, they have a rich topological structure.

\item[(ii)]

Partial \xas\ of functions, such as the scale of the $L^p$ spaces on [0,1] or the
lattice generated by the family $\{  L^p(\BR), \, 1 \leqslant p \leqslant \infty \}$.
These  \pa s have the peculiarity of carrying two structures: they are simultaneously
a partial inner product space ({PIP}-space) \cite{PIP12}-\cite{PIP4} and an abelian \pa\
under pointwise multiplication --- and the two structures fit perfectly. We will say
more about this class in Section 4 below.

\item[(iii)]

Partial \xas\ of operators, such as sets of operators on a { PIP}-space or \po s. Here the algebraic structure is richer and we will have only partial results (see Section 5).
\eni

\noindent
This discussion at the same time suggests the organization of the paper. First we
start from an abstract \pa, focusing on the structure of its multiplier spaces, as
described in \cite{ag}, this is Section 2.
In Section 3 we propose a definition of \tpa, based on the multiplier structure
(which embodies all the information about the partial multiplication), and check
that it applies indeed to the simple cases mentioned under (i) above.
Sections 4 and 5 contain a full discussion of the cases (ii) and (iii), respectively.
In addition to the $L^p$ spaces, we will also consider in Section 4 a wide class  of generalizations, the so-called {\em amalgam spaces} introduced by N. Wiener \cite{wiener}.

This paper by no means pretends to exhaust the subject. On the contrary, it is more a
program, with many questions remaining open. Yet we feel the proposed definition is
natural, in the sense that, in the examples mentioned, it brings almost perfect
coincidence between the algebraic (multiplier) structure and the topological one. Only  applications will tell whether our definition has to be made
more (or less) restrictive.


\section{\hspace{-5mm}. Spaces of multipliers on \pa s}

For the sake of completeness, we recall first the basic definitions. A 
 {\em \pa} is a complex vector
space \A, endowed with an  involution $x \mapsto x\ha$ 
(that is, a bijection such that $x\haa = x$)
and a partial multiplication defined by a set $\Gamma \subset \A \times \A$
(a binary relation) such that:

$\;$(i) $(x,y) \in \Gamma$ implies  $(y\ha,x\ha) \in \Gamma$;

$\,$(ii)    $(x,y_1), (x,y_2) \in \Gamma$ implies
$(x, \lambda y_1 + \mu  y_2) \in \Gamma, \, \forall \,\lambda,\mu \in \BC; $

(iii) for any $(x,y) \in \Gamma$, there is defined a product
$x \cdot y \in \A$, which is distributive w.r. \\ 
\hspace*{13mm} to the addition.

\noindent
Notice that the  partial multiplication is {\em not} required to be associative (and
often it is not). We shall assume the \pa\ \A\ contains a unit $e$, i.e.
$e\ha = e, \, (e,x) \in \Gamma, \, \forall \, x \in \A$,
and $ e \cdot x = x \cdot e = x, \, \forall \, x \in \A$. (If \A\ has no unit, it may
always be embedded into a larger \pa\ with unit, in the standard fashion 
\cite{am}.)

Given the defining set $\Gamma$, spaces of multipliers are defined in the obvious way:
\beano
(x,y) \in \Gamma &\Longleftrightarrow& 
      x \in L(y) \, \mbox{ or $x$ is a left multiplier of $y$ } \\
&\Longleftrightarrow& 
      y\in R(x) \, \mbox{ or $y$ is a right multiplier of $x$} .
\enano
For any subset $\N \subset \A$, we write
$$
L\N =  \bigcap_{x \in \n} L(x), \quad R\N =  \bigcap_{x \in \n} R(x),
$$
and, of course, the involution exchanges the two:
$$
(L\N)\ha = R\N\ha, \quad (R\N)\ha = L\N\ha.
$$
Clearly all these multiplier spaces are vector subspaces of \A, containing $e$. 

The \pa\ is {\em abelian} if $L(x) = R(x),\, \forall \, x \in \A$, and then 
$ x \cdot y = y \cdot x, \, \forall \, x \in L(y)$. In that case, we write simply
for the multiplier spaces 
$L(x) = R(x)  \equiv  M(x), \; L\N = R\N \equiv M\N \; (\N \subset \A)$.

Now the crucial fact is that the couple of maps $(L,R)$ defines a {\em Galois
connection}  on the complete lattice of all vector subspaces of \A\ (ordered by
inclusion), which means that 
(i)  both $L$ and $R$ reverse order;
and
(ii) both $LR$ and $RL$ are closures, i.e.:
$$
\begin{array}{lll}
\N \subset LR\N  \,&\mbox{ and }\,&LRL = L
\medskip\\
\N \subset RL\N  \,&\mbox{ and }\, &RLR = R.
\end{array}
$$
Let us denote by $\F^L$, resp. $\F^R$, the set of all $LR$-closed, resp. 
$RL$-closed, subspaces of \A:
\beano
\F^L = \{ \N \subset \A : \N = LR\N \}, \\
\F^R = \{ \N \subset \A : \N = RL\N \}.
\enano
both ordered by inclusion. Then standard results from universal algebra yield the full
multiplier structure of \A\ \cite{ag,ak}:


\betheo
 --
Let \A\ be a \pa\ and $\F^L$, resp. $\F^R$, the set of all $LR$-closed, resp. 
$RL$-closed, subspaces of \A, both ordered by inclusion. Then

(1) $\F^L$ is a complete lattice with lattice operations
$$
\M \wedge \N = \M \cap \N , \quad \M \vee \N = LR(\M + \N).
$$
 The largest element is \A, the smallest $L\A$. 

(2) $\F^R$ is a complete lattice with lattice operations
$$
\M \wedge \N = \M \cap \N , \quad \M \vee \N = RL(\M + \N).
$$
The largest element is \A, the smallest $R\A$. 

(3) Both $L: \F^R \to \F^L$ and $R: \F^L \to \F^R$ are lattice anti-isomorphisms:
$$
L(\M \wedge \N) = L\M \vee L\N, \,\mbox{ etc.},
$$

(4) The involution $\N \leftrightarrow \N\ha$ is a lattice isomorphism between 
$\F^L$ and $\F^R$.
                \hfill \mbox{$\square$}
\entheo

\noindent
In addition to the two lattices $\F^L$ and $\F^R$, it is useful to consider the
subset
$\F^\Gamma \subset \F^L \times \F^R$ consisting of {\em matching pairs}, that is:
$$
\F^\Gamma  = \{ (\N,\M) \in \F^L \times \F^R :  \N = L\M 
 \,\mbox{ and }\,  \M = R\N \}.
$$
Indeed these pairs describe completely the partial multiplication of \A, for we can
write:
$$
(x,y) \in \Gamma \; \Longleftrightarrow \; 
\exists \, (\N,\M) \in \F^\Gamma \,\mbox{ such that }\, x \in \N , y \in \M.
$$

\section{\hspace{-5mm}. Topological \pa s: Definition and first
\protect \\  examples}

Let \A\ be a \pa\ with unit and assume it carries a locally convex, Hausdorff, topology $\tau$,
which makes it
into a locally convex topological vector space $\A[\tau]$ (that is, the vector space
operations are $\tau$-continuous). We  denote by $\{ p_\alpha \}$ a
(directed) set of seminorms defining $\tau$.

As we saw in Section 2, the \pa ic structure of \A\ is completely characterized by
its spaces of left, resp. right, multipliers. Thus, quite naturally, we describe
the topological structure of  $\A[\tau]$ by providing all 
spaces of  multipliers with appropriate topologies.

We start with the following observation. Let $\M \in \F^R$. To every $x \in L\M$, one may associate a linear map $T^L_x$ from \M\ into \A:
$$
T^L_x(a) = xa, \quad a \in \M, \; x \in L\M.
$$
This allows to define the topology $\rho_{\m}$ on \M\ as the weakest 
locally convex topology on \M\ such that all maps $T^L_x, \,x \in L\M, $ are continuous
from \M\ into $\A[\tau]$. This is of course a projective topology. 
In the same way, the topology $\la_{\n}$ on $\N \in \F^L$ is the  weakest 
locally convex topology on \N\ such that all maps 
$T^R_y: a \mapsto ay, \,y \in R\N, $ are continuous from \N\ into $\A[\tau]$.

In terms of the seminorms  $\{ p_\alpha \}$ defining $\tau$, it is clear that the
topology   $\rho_{\m}$ on \M\ may be defined by the seminorms
$$
p^x_{\alpha,\rho}(a) = p_{\alpha}(xa), \; x \in L\M,
$$
and the topology $\la_{\n}$ on \N\ by the seminorms
$$
p^y_{\alpha,\la}(a) = p_{\alpha}(ay), \; y \in R\N.
$$
It follows immediately from the definition that, whenever $\M_1, \M_2  \in \F^R$
are such that  \mbox{$\M_1 \subset \M_2$,} then the topology   $\rho_{\m_1}$
is finer than the topology $(\rho_{\m_2}  \up \M_1)$ induced by $\M_2$
 on $\M_1$.
In other words, the embedding $\M_1 \to \M_2$ is a {\em continuous injection}.

Take now \A\ itself. It carries three topologies,  $\tau, \,\rho_{\ga}$ and
$\la_{\ga}$. How do they compare? The topology  $\rho_{\ga}$ makes
all maps
$$
T^L_x: a  \mapsto   xa, \quad a \in \A, \; x \in L\A
$$
continuous. This is true in particular for  $T^L_e$, where  $e$ is the unit, 
which means precisely that the embedding $\A[\rho_{\ga}] \to \A[\tau]$
is continuous. The same applies of course to $\A[\la_{\ga}] \to \A[\tau]$.
In other words, both $\rho_{\ga}$ and $\la_{\ga}$ are finer than $\tau$.

As a consequence, since $\tau$ was assumed to be Hausdorff, all topologies
$\rho_{\m}, \, \M \in \F^R$ and $\la_{\n}, \, \N \in \F^L$, are Hausdorff.

Now, for reasons of coherence, it would be preferable that all 
three topologies on \A,  $\tau, \,\rho_{\ga}$ and
$\la_{\ga}$ be equivalent. Here is a handy criterion.


\belem
--
Let $\A[\tau]$ be a \pa\ with locally convex  topology $\tau$. Then:

(1) The projective topology $\rho_{\ga}$ on \A\ is equivalent to $\tau$
iff, for each $x \in L\A,$ the map $T^L_x: a  \mapsto   xa$ is continuous
from $\A[\tau]$ into itself.

(2) The projective topology $\la_{\ga}$ on \A\ is equivalent to $\tau$
iff, for each $y \in R\A$, the map $T^R_y: a  \mapsto   ay$ is continuous
from $\A[\tau]$ into itself.
\label{cont}
\enlem

\proof 
  (1) We know that $\rho_{\ga} > \tau$. Since $\rho_{\ga}$ is by definition the weakest
topology on \A\ that makes the map $T^L_x$ continuous, the statement follows.

(2) Same argument.
\enproof

\noindent
Assume now that the involution $x \mapsto x\ha$ is continuous in $\A[\tau]$.
If $\M \in \F^R$ and $a \in \M$, then $a\ha \in \M\ha$, by Theorem 2.1 (4). Then,
for     $x \in R\M\ha$ and every seminorm $p_{\alpha}$ of $\A[\tau]$,
there is a seminorm $p_{\beta}$ such that, for some positive constant $c$, 
$$
p^x_{\alpha,\la}(a\ha) = p_{\alpha}(a\ha x) = p_{\alpha}( (x\ha a)\ha)
\leqslant c \, p_{\beta}(x\ha a) = c \,p^{x\ha}_{\beta,\rho}( a).
$$ 
Similarly, if $\M = \M\ha \in \F^L \cap \F^R, \, a = a\ha \in \M$ and
$x \in L\M$, we get
$$
p^x_{\alpha,\rho}(a\ha) = p_{\alpha}(xa) = p_{\alpha}( (a x\ha )\ha)
\leqslant c \, p_{\beta}(a x\ha ) = c \,p^{x^*}_{\beta,\la}(a).
$$
Thus we have proven


\belem
--
Let $\A[\tau]$ be a \pa\ with locally convex  topology $\tau$. 
Assume that the involution $x \mapsto x\ha$ is $\tau$-continuous. Then:

(1) For every $\M \in \F^R$, the involution is continuous from
$\M[\rho_{\m}]$ into $\M\ha[\la_{\m\ha}] \in \F^L$.

(2) Let   $\M = \M\ha \in \F^L \cap \F^R$. Then the topology $\rho_{\m}$ is equivalent
to $\la_{\m\ha} = \la_{\m}$ on self-adjoint elements of \M.\enproof
\enlem
According to our goal to make the algebraic and the topological structure coincide as much as
possible, on a topological \pa, we will naturally require that all three topologies
$\rho_a$, $\lambda _a$ and $\tau$ coincide and that the involution be continuous.  Let us now look
at multiplier spaces $\M \in \F^R$.  If $\M_1 \subset \M_2$, we have seen that the embedding is
continuous.  In order to make the structure tighter, we should also require that $\M_1$ be {\em
dense} in $\M_2[\rho_{\m_2}]$.  This is true in many examples, typically the function spaces of
Section 4 (such a condition is of course reminiscent of { PIP}-spaces --- which these function spaces
actually are also).  Of course it is enough to require that $R\A $ be dense in each
$\M[\rho_{\m}] \in \F^R$.  Indeed, if $R\A \subset \M_1 \subset\M_2$, and $R\A$ is dense in $\M_2$
for $\tau_{\m_2}$, so is a fortiori $\M_1$.  But this condition is still too strong (and hardly
verifiable in practice, because $\F^R$ is too large).  To go beyond, we introduce the notion of
{\em generating family}, a notion equivalent to that of {\em rich subset} for a compatibility
relation, as described in \cite{PIP3}.

\beee
--
A subset $\I^R$ of $\F^R$ is called a {\em generating family} if

(i) $R\A \in \I^R$ and  $\A \in \I^R$.

(ii) $x \in L(y)$  iff $\,\exists \,\M \in \I^R$  s.t. $y \in \M, x \in L\M$.
\\
A generating family for $\F^L$ or $\F^\Gamma$ is defined in a similar way.
\enee

Clearly, if $\I^R$ is a generating family for $\F^R$,
$\I^L = L \I^R = \{L\M : \M \in \I^R \}$ is generating for $\F^L$ and $\I^\Gamma = \I^L \times \I^R$ is generating for
$\F^\Gamma$.

The usefulness of this notion is twofold :
\bei
\item[(i)] 
if $\I^R$ is generating for $\F^R$, so is the sublattice $\J^R$  of $\F^R$ generated 
from $\I^R$ by {\em finite} lattice operations $\vee $ and $\wedge$. 
\item[(ii)] 
if $\I^R$ is generating, the {\em complete} lattice generated by $\I^R$ is $\F^R$ itself.
\eni
We make immediate use of this last property for weakening the density condition.


\belem
--
Let $\A[\tau]$ be a partial *-algebra with topology $\tau$.  Assume there exists a generating family
$\I^R$ for $\F^R$ such that $R\A$ is dense in $\M[\rho_{\m}]$ for every $\M \in \I^R$.  Then, for
any pair $\M_1, \M_2 \in \F^R$ such that $\M_1 \subset \M_2,\, \M_1$ is dense in $\M_2
[\rho_{\m_2}]$.
\enlem

\proof  Let $ \M \in \F^R$.  Since $\F^R$ is the lattice completion of $\I^R$, we may write 
$$
\M =  \bigcap_\alpha \N_\alpha, \; \N_\alpha \in \I^R, \; \N_\alpha \supset \M.
$$
By assumption, $R\A$ is dense in every $\N_\alpha [\rho_{\n_\alpha}]$.  Then it is also dense in
their intersection, endowed with the projective  topology, since the latter is the projective
limit of a directed set of subspaces \cite{schaefer}. But this is precisely $\M [\rho_{\m}]$.

Let now $\M_1 \subset \M_2$, both in $\F^R$.  Since $R\A$ is dense in $\M_2 [\rho_{\m_2}]$, so is
$\M_1$.
\enproof

Putting all these considerations together, we may now state our definition of topological partial
*-algebra.


\beee
\label{def3.5} --
Let $\A [\tau]$ be a \pa, which is a TVS for the locally convex topology $\tau$.  Then
$\A [\tau]$ is called a {\em  \tpa} if the following two conditions are
satisfied :

\bei
\item[(i)]
 the involution $a \mapsto a\x$ is $\tau$-continuous;

\item[(ii)]
 the maps $a \mapsto xa$ and  $a \mapsto ay$ are $\tau$-continuous for all $x
\in L\A$ and $y \in R\A$.
\eni
The \tpa\ $\A [\tau]$ is said to be {\em tight}, if, in  addition,
\bei
\item[(iii)]
 there is a generating family $\J^R$ for $\F^R$ such that $R\A$ is dense in
$\M[\rho_{\m}]$ for each $\M \in \J^R$.
\eni
\enee
As we shall see in the following sections, these conditions will be satisfied in most examples we
consider.  But before that, it is worth considering again the density condition (iii).  According
to Lemma 3.4, its effect is to ensure that all the embeddings $\M_1 \subset \M_2 \;(\M_1, \M_2
\subset \F^R)$ have dense range.  An equivalent statement would be that the dual of 
$\M_2 [\rho_{\m_2}]$ be a subspace of the dual of $\M_1 [\rho_{\m_1}]$.  Thus we characterize these
dual spaces.

\belem
--
Let $\M \in \F^R$, with its projective topology $\rho_{\m}$.  Then a linear functional $F$ on
$\M$ is $\rho_{\m}$-continuous if and only if it may be represented as 
\be
 F(x) = \sum ^n _{i=1} G_i (a_i x),
\label{form}
\en
 where each $G_i$ is a $\tau$-continuous functional on \A\
and $a_i \in L\M$,  i = 1 \ldots n.
\label{form1}
\enlem

\proof  
If $G$ is $\tau$-continuous and $a \in L\M$, we get
$$
|G(a x)| \leqslant p(a x),
$$
where $p$ is a continuous seminorm on $\A[\tau]$. It is clear that $p^a(x) \equiv p(a x)$ is 
a continuous seminorm on $\M[\rho_{\m}]$. Therefore, $ F(x) = \sum ^n _{i=1} G_i (a_i x), \, x \in
\M,$ is $\rho_{\m}$-continuous for $G_i$ and $a_i$ satisfying the assumptions.

Conversely, let $F$ be $\rho_{\m}$-continuous on \M. Then there exist seminorms 
$p_{\alpha_1}, \ldots, p_{\alpha_n}$, $ a_1, \ldots, a_n \in L\M$ and $c>0$ such that
$$
 |F(x)| \leqslant c \sum ^n _{i=1}p_{\alpha_i} (a_i x), \; x \in \M.
$$
Let us consider the following subspace \gK\ of $\A \oplus \A \ldots \oplus \A$ ($n$ terms):
$$
\gK = \{ (a_1 x, a_2 x, \ldots, a_n x) | x \in \M \}.
$$
Then the functional $G((a_1 x, a_2 x, \ldots, a_n x) ) = F(x)$ is linear and continuous on \gK\ with
respect to the product topology defined by  $\tau$. By the Hahn-Banach theorem, $G$ can be
extended to a continuous linear functional on $\A \oplus \A \ldots \oplus \A$ ($n$ terms).
This implies that there exist linear functionals $G_i$ on \A\ such that
$G((Y_1, \ldots, Y_n)) = \sum ^n _{i=1} G_i(Y_i)$. Therefore we conclude that 
$ F(x) = \sum ^n _{i=1} G_i (a_i x)$.
\enproof

It is instructive to rewrite the form (\ref{form}) in terms of tensor products :
$$
F = \sum ^n _{i=1} G_i \otimes a_i,  \;G_i \in \A',  \, a_i \in L\M.
$$
Then the statement of Lemma 3.6 may be reformulated as:
$$
 \M'  =  \A' \otimes L\M\ /\ \K [\M'].
$$
where the kernel $\K[\M']$  consists of the forms in $ \A' \otimes L\M$ that vanish on \M\ :
$$
\K[\M'] = \{ \sum^n _1 G_i \otimes a_i \in \A' \otimes L\M :
  (G_i \otimes a_i) (x) = 0, \forall \, x \in \M \}.
$$
In this language, condition (iii) in Definition \ref{def3.5} says  a sufficient condition for the
embedding 
$\M_1\subset \M_2$ to have dense range is that 
\be 
\label{kern}
 \K[\M'] = \K[(R\A)'] \cap (\A' \otimes L\M), \; \forall \, \M \in \J^R.
\en
In other words, an element of $\A' \otimes L\M$ vanishes on $\M$ iff it vanishes on $R\A$, which of
course amounts to say that $\M'$ is a subspace of $(R\A)'$.  To see what can happen, it is amusing to
consider the extreme case where $R\A$ is one-dimensional, ie. 
$R\A = {\mathbb C} e$.  Then indeed one
sees easily that
$\K[\A'] = \{0\}$, whereas $\K[(R\A)']$ is of codimension 1, and thus making $\M \equiv \A$ in
(\ref{kern}),
$
\K[\A'] \subsetneqq \K [(R\A)']\cap \A'.$

The previous discussion is summarized by the following

\beprop 
--
Let $\A [\tau]$ be a topological \pa\ and $\J^R$ a generating
family for $\F^R$. If the dual of each $\M[\rho_{\m}]$ can be
identified with a subspace of $(R\A [\rho_{R\ga}])'$, then $\A
[\tau]$ is a tight topological \pa.  
\enprop
Actually, the tightness condition, despite its appearance, is familiar in functional analysis. As we shall see in Section 4, many families of function spaces (such as $L^p$ spaces, Sobolev spaces, etc.) can be recast into 
\tpa s. Tightness, in these examples, simply expresses the existence of a space of universal multipliers which is dense in each one of the spaces of the family. This is often realized by suitable classes of $C^\infty$ functions.

As discussed in the Introduction, we feel that Definition \ref{def3.5} is natural, in the sense that
it forces the topological structure determined by $\tau$ to be consistent with the multiplier
structure of $\A$. 
\\ As an illustration of the developed ideas, we consider  two abstract examples.

\subsection{Topological quasi *-algebras}
Let $(\A,\A_o)$ be a topological quasi-algebra, that is, $\A_o$ is a topological \xa\ such that the
multiplication is separately, but not jointly, continuous for the topology of $\A_o$ and the latter
is not complete, and $\A$ is the completion of $\A_o$.
Thus $\A$ is only a \pa: the product $xy$ is defined only if either $x$ or $y$
belongs to $\A_o$.
Clearly, $(\A,\A_o)$ is a (trivial) \pa\ with $L\M = R\M = \A_o$ and $\A_o$ is
dense in $\A$.   Every topological quasi *-algebra is a tight \tpa. 
We remark that, according to the previous discussion, $\A_o$ becomes in natural way a topological *-algebra with respect to the topology defined by the seminorms:
$$
p^x_{\alpha}(a) = max\{p_{\alpha}(xa), p_{\alpha}(ax)\}, \; x \in \A,
$$
where the $p_\alpha$'s are the seminorms defining the topology $\tau$ of $\A$. This topology is finer than the initial topology of $A_o$.

\subsection{\em CQ\xa s}
This family of \pa s appears, under several aspects, the natural extension of C\xa s in the partial algebraic setting.
The definition of CQ\xa\ that we will give here is different from the original one \cite{bagtrap1,bagtrap2}, but fully equivalent. 

\begin{defin}
Let $\A$ be a right Banach module over the C*-algebra $\A_\flat$, with isometric involution $*$ and such that $\A_\flat \subset \A$. We say that $\{\A, *, \A_\flat, \flat\}$ is a CQ*-algebra if
\begin{itemize}
\item[(i)] $\A_\flat$ is dense in $\A$ with respect to its norm $\|\,\|$
\item[(ii)]$\A_o:=\A_\flat \cap \A_\sharp$ is dense in $\A_\flat$ with respect to its norm $\|\,\|_\flat$
\item[(iii)]$\|B\|_\flat = \sup_{A \in \A}\|AB\|,\quad B \in \A_\flat$ 
\end{itemize}
\end{defin}
The following picture can be of some help in understanding the situation: 

\vspace{1cm}
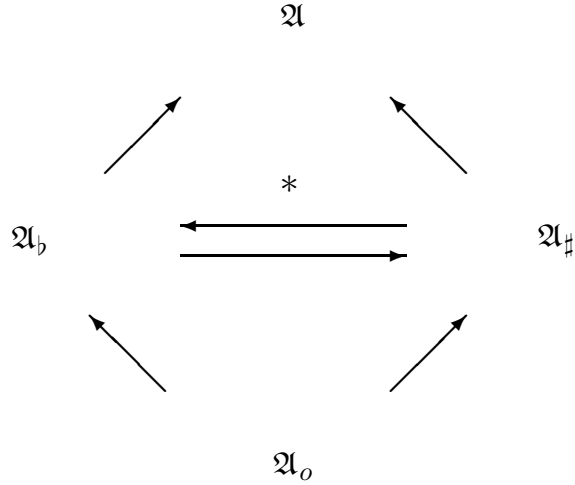
\begin{center}
\setlength{\unitlength}{1mm}
\begin{picture}(80,80)
\thicklines
\put(35,60){\makebox(10,10){\large $\A$}}
\put(15,44){\vector(1,1){10}}
\put(63,44){\vector(-1,1){10}}
\put(23,15){\vector(-1,1){10}}
\put(53,15){\vector(1,1){10}}
\put(25,33){\vector(1,0){30}}
\put(55,37){\vector(-1,0){30}}
\put(37,40){\makebox(5,5){\large $*$}}
\put(0,30){\makebox(10,10){\large $\A_\flat$}}
\put(70,30){\makebox(10,10){\large $\A_\sharp$}}
\put(35,0){\makebox(10,10){\large $\A_o$}}
\end{picture}
\end{center}

It is clear from the above definition that a CQ\xa\ is automatically a  tight
\tpa.

To give the flavor of this construction, consider the following simple example \cite{bagtrap4}.
  Take a  (Gel'fand) triplet of Hilbert spaces  
\be
\H_\lambda \; \subset \; \H \; \subset \; \H_{\bar\lambda},
\en
where $\H_\lambda$ is, for instance, the domain of some self-adjoint operator $H>0$ 
(such that $(1+H)^{-1/2}$ is Hilbert-Schmidt)  with the graph norm
$\| (1+H)^{1/2}f \|$, and $\H_{\bar\lambda}$ is the anti-dual of
$\H_\lambda$ with respect to the inner product of $\H$ (i.e. the
norm on $\H_{\bar\lambda}$ is $\| (1+H)^{-1/2}f \|$).  Then, if one
makes the following identifications:

. $\A\ = \B(\H_\lambda,\H_{\bar\lambda})$, a Banach space;

. $R\A\ = \B(\H_\lambda)$, a C\xa;

. $L\A\ = \B(\H_{\bar\lambda})$, also a C\xa;

. $\A_o = \B(\H_\lambda) \cap \B(\H_{\bar\lambda}) = \{ A \in \B(\H_\lambda,\H_{\bar\lambda}):
A \; {\rm and } \;A\x \in \B(\H_\lambda) \},$
\\
one can show that $\B(\H_\lambda,\H_{\bar\lambda})$ is a CQ\xa\ and a tight \tpa.

Due to its definition, a CQ\xa\ turns out to be useful in the description of certain quantum models in many cases where, for some physical reason, $R\A$ is not large enough to include all the relevant observables of the given physical model, with their time-evolutes \cite{bagtraph}.\\
As for the structure, a CQ\xa\ can be viewed as the completion of a C\xa\ with respect to a weaker norm: this is exactly the case of {\em proper} CQ\xa s ($R\A=L\A;\, \sharp=\flat$) \cite{bagtrap1} or even, under stronger assumptions, in the non-proper case \cite{bagintrap}\\
Of particular interest is the case of a {\em *-semisimple} CQ\xa\ (i.e., with trivilal *-radical). In this case, the analogy with C\xa s becomes closer and closer.\\
First, for *-semisimple CQ\xa s, it is possible to define a {\em refinement} of its partial multiplication: in this way, its lattices of multipliers become absolutely non-trivial. This allows an extension of certain facts of the familiar functional calculus for C\xa s.\\
Second, the abelian case is completely understood: an abelian *-semisimple CQ\xa\ can be realized as a CQ\xa\ of functions by means of a generalized Gel'fand transform.\\
All these facts are discussed in \cite{bagtrap2, bagtrap3}.\\
For all these reasons we consider CQ\xa s, as a first step toward a more general study of {\em partial C\xa s} which is still to be carried out.
     
\vspace{4mm}  
In the following two sections, we shall discuss in detail more sophisticated examples, namely
functions spaces that will yield abelian topological partial *-algebra (Section 4) and partial
*-algebras of operators (Section 5).

\section{\hspace{-5mm}. Examples I : Topological \pa s of functions}

\subsection{\hspace{-4mm}. $L^p$ spaces on a finite interval}

A standard example of an abelian partial *-algebra \cite{ait1}
is the space $L^1([0,1],dx)$, equipped with the partial multiplication:
\be
f \in M(g) \; \Leftrightarrow \; \exists \, q \in [1, \infty] \; \mbox{ such that }
f \in L^q, \, g \in L^{\bar q}, \;1/q + 1/{\bar q} = 1.
\label{mult}
\en
A similar structure may be given for every $L^p$.  In fact one can show \cite{bagtrap3} that
every space $L^p(X,d\mu)$, with  $X$ a compact space and $\mu$ a Borel measure on $X$, is an abelian
CQ*-algebra, with $\A_o = \C(X)$, the space of continuous functions.

What we envisage here  is the chain of all spaces $L^p$ at once, and for simplicity we take for
$(X,\mu)$  the interval [0,1] with Lebesgue measure.  Thus we consider the chain 
$\I= \{ L^p([0,1],dx),  1 \leqslant p \leqslant \infty\}$, with $ L^p \subset L^q, \, p > q$.
  For $1 < p < \infty$, every
space $L^p$ is a reflexive Banach space with dual $ L^{\bar p} \;(1/p + 1/{\bar p} = 1)$. 
Notice that duality in the sense of Banach spaces coincides with duality  for the inner product  of
$L^2$ thanks to H\"older's inequality. 

Now, being a chain, $\I$ is of course a lattice, albeit not a complete one.  The lattice
completion of $\I$, denoted $\F$, may be characterized explicity from the work of Davis et al.
\cite{dav} (see also \cite{PIP3,ak=refin}).
Define the two spaces : 
$$
L^{p-} = \bigcap_{1 \leqslant  q<p}  L^q, \quad L^{p+} = \bigcup_{p<q\leqslant \infty}  L^q.
$$
Then for $1<p\leqslant \infty$, $L^{p-}$,  with the projective topology, is a non-normable reflexive
Fr\'echet space, with dual
$L^{\bar p +}$.  And for $1 \leqslant p < \infty, \; L^{p+}$,   with the inductive topology, is a
nonmetrizable complete DF-space, with dual $L^{\bar p -}$ (a DF-space is the dual of a Fr\'echet
space, necessarily non metrizable, unless the space and its dual are both Banach spaces
\cite{schaefer}).  Finally the following inclusions are strict:
\be
L^{p+} \;\subset \; L^{p} \;\subset \; L^{p-} \; 
\subset \; L^{q+}  \quad (1<q<p<\infty),
\label{Lptriplet}
\en 
all embeddings in (\ref{Lptriplet})  are continuous and have dense range.
Then the complete lattice $\F$ generated by $\I$ is also a chain, obtained by replacing each   
$L^{p} \, (1<p<\infty)$ by the corresponding triplet as in (\ref{Lptriplet}) and adding the two
spaces $L^\omega \equiv L^{\infty-}$ (the so-called Arens space) and  $L^{1+}$:
$$
L^{\infty}\;\subset \;L^\omega \;\subset \; \ldots \;\subset \; L^{p+} \;\subset \; L^{p}\;\subset
\; L^{p-}\;\subset \;\ldots \;\subset \;L^{1+}\;\subset \;L^{1}.
$$
Of course it would be more natural to index the spaces by $1/p$, but traditions are respectable!
Thus we take systematically our chains of spaces as increasing to the right, with decreasing $p$.

Now we turn to the partial *-algebra structure.  The partial multiplication on the space
$L^1([0,1],dx)$  is defined as in (\ref{mult}),
i.e. $\I$ is a generating family. For  computing multiplier spaces, define the following set, which
characterizes the behavior of an individual vector $f \in L^1$:
$$
J(f) = \{ q \geqslant 1 : \, f \in L^q \}
$$
and let $p = \sup J(f)$, with $1 \leqslant   p \leqslant \infty$.

\setlength{\unitlength}{1cm}

\begin{center}
\begin{picture}(6,2)
\thicklines
\put(5,1.4){\makebox(0,0){$J(f)$}}
\put(0,0.6){\makebox(0,0){$\infty$}}
\put(0,1.05){\makebox(0,0){$\shortmid$}}

\put(3,0.6){\makebox(0,0){$ p$}}
\put(3,1.05){\makebox(0,0){$\shortmid$}}

\put(7,0.6){\makebox(0,0){1}}
\put(7,1.05){\makebox(0,0){$\shortmid$}}

\put(0,1){\line(1,0){3}}
\put(3,1.05){\line(1,0){4}}
\put(3,1){\line(1,0){4}}
\end{picture}

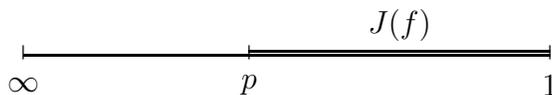
Figure 1 : The set $J(f)$.
\end{center}
\bigskip

\noindent
We distinguish two cases:

(i) $ J(f) = [1, p]$,   a closed interval, i.e.
$f \in L^{  p}$, but $f \not\in L^{s}, \forall \, s >  p.$
Then it is easily seen  that $M(f) = L^{\bar p}$ \cite{bagtrap3}.

(ii) $ J(f) = [1, p)$,   a semi-open interval, i.e.
$f \in L^{q}, \, \forall \,  q <   p $, hence 
$f \in L^{p-}= \bigcap_{ q < p}  L^{ q}$, but $f \not\in L^{p}$. Then
$M(f) = \bigcup_{\bar q > \bar p} L^{\bar p} = L^{\bar p+}.$

From these results, it follows immediately that 
 $$
ML^{ p} = L^{\bar p},\quad  ML^{p-} = L^{\bar p+},\quad ML^{ p+} = L^{\bar p-}.
$$
Notice that, if we define $f,g$ to be multiplicable whenever $fg \in L^1$, then the space of
multipliers $M(f)$ of a given element $f$ is more complicated, but we still have  
$ML^{p} = L^{\bar p},$ etc, as follows from \cite{bagtrap3}.

 As for the multiplier topologies, we also have that 

. $\rho_{L^p}$ is the $L^p$ norm topology

. $\rho_{L^{p-}}$ is the Fr\'echet projective topology on $L^{p-}$

. $\rho_{L^{p+}}$ is the DF topology on $L^{p+}$.

\noindent
For both $\I$ and $\F$, the smallest space  is $L^\infty = ML^1$, and it is dense in all the
other ones. The involution $f \mapsto \bar f$ is of course $L^1$-continuous. The multiplication is
continuous from $L^\infty \times L^1$ into $L^1$.  In fact it is not only separately, but even
jointly continuous and similarly from $L^p  \times L^{\bar p}$ and from
$L^{p-}  \times L^{\bar p+}$ into $L^1$, thanks to H\"older's inequality and the fact that all
topologies are either Fr\'echet or DF \cite{schaefer}.
Since this result is general, we state it as a proposition.


\beprop
--
Let $\A[\tau]$ be a \pa\ with locally convex  topology $\tau$, and $\I^R$ a generating family.
Assume that:

(1) $\tau$ is a norm topology and $\A[\tau]$ is a Banach space.

(2)  Each space $\N \in \I^L$, with the topology $\la_{\n}$, and each 
 space $\M \in \I^R$, with the topology $\rho_{\m}$, is a Banach space. 

Then the multiplication is jointly continuous from $L\M \times \M$ into \A\, for every $\M
\in \I^R$, and one has
\be
\label{banach}
\| a b\| \leqslant \| a \|_{L\m} \, \| b \|_{\m}, \; \forall \, a \in L\M, \; b \in \M.
\en
\enproof
\enprop
The proof of (\ref{banach}) essentially reduces to the principle of uniform boundedness. Indeed, for fixed $a \in L\M$, the map $a \mapsto T_a^L$
is continuous from $L\M$ into the space of bounded operators on $\M$, which gives $\| a b\| \leqslant c \, \| a \|_{L\m} \, \| b \|_{\m}$ for some constant $c>0$. The latter may then be eliminated by renormalizing all norms by a factor $c$. Notice that (\ref{banach}) is strongly reminiscent of a H\"older condition. In fact it reduces to the latter in the case of  $L^p$ considered as a \tpa, as discussed in Section 4 below.

A \pa\ that satisfies the conditions of Proposition 4.1 may be called a {\em Banach \pa,} since the relation (\ref{banach}) is the analogue of the characteristic property of Banach algebras
A similar result holds if one of the spaces $\M, \; L\M$ is a Fr\'echet space and the other a
DF-space, with $\A[\tau]$ itself a Fr\'echet space.

In conclusion, the topological structure, the { PIP}-space structure and the multiplier structure of
$\I$ all coincide, and we have a tight \tpa.

By the same token, we can consider every space $L^p$, as a topological \pa, simply by replacing the
partial multiplication (\ref{mult}) by the following one:
\be
f \in M(g) \; \Leftrightarrow \; \exists \, r, s \in [p, \infty],  \;1/r + 1/s = 1/p, \; \mbox{ such
that } f \in L^r, \, g \in L^{s}.
\label{mult_p}
\en
This amounts exactly to replace \I\ or \F\ by the (complete) sublattice indexed by $[p,\infty]$.
The rest is identical. 
\bigskip

\subsection{\hspace{-4mm}. The spaces $L^p(\BR,dx)$  } 

We turn now to the $L^p$ spaces on $\BR$. If we consider the family
$\{L^p(\BR) \cap L^1(\BR), \, 1 \leqslant p  \leqslant \infty\}$, we obtain a scale  similar to the previous one (except that the individual spaces are not complete), which may be used to endow $L^1(\BR)$ with the structure of a tight \tpa.

However, the spaces $L^p(\BR)$ themselves no longer form a
chain, no two of them being comparable. We have only 
$$
 L^p \cap L^q \subset L^s, \, \forall \, s \; \mbox{ such that } \; p<s<q.
$$
Hence we have to take the lattice generated by
$\I= \{ L^p(\BR,dx), 1 \leqslant p \leqslant \infty\}$, that we call   \J. 
The extreme spaces of the lattice are, respectively:
$$ 
V_J^\# = \bigcap_{1\leqslant q \leqslant \infty} L^q, \quad \mbox{ and } \quad V_J = \bigcup_{1\leqslant q \leqslant \infty} L^q
= \sum_{1\leqslant q \leqslant \infty} L^q.
$$
Here too, the lattice structure allows to give to $V_J$ a structure of \tpa, as we shall see now.

The lattice operations on \J\ are easily described
\cite{PIP3,ak=refin,berghlof,dav}: 
\bei
\item
$L^p \wedge L^q = L^p \cap L^q$ is a Banach space, with the projective (topology corresponding to
the) norm 
$$
\| f \|_{p \wedge q} = \| f \|_p + \| f \|_q.
$$
\item
$L^p \vee L^q = L^p + L^q$ is a Banach space, with the inductive (topology corresponding to
the) norm 
$$
\| f \|_{p \vee q} = \inf_{f=g+h}\left(\| g \|_p + \| h \|_q\right), \; g \in L^p, \, h \in L^q.
$$
\item
For $1<p,q<\infty$, both spaces $L^p \wedge L^q$ and $L^p \vee L^q$ are reflexive and
$(L^p \wedge L^q)' = L^{\bar p} \vee L^{\bar q}$.
\eni

At this stage, it is convenient to introduce a unified notation:
$$
L^{(p,q)} = \left\{ \begin{array}{ll}
L^p \wedge L^q, & \mbox{ if } \; p \geqslant q, \\
L^p \vee L^q, & \mbox{ if } \; p \leqslant q.
\end{array}
\right.
$$
Thus, for $1 < p,q < \infty$, each space $L^{(p,q)}$ is a reflexive Banach space, with dual 
  $L^{(\bar p,\bar q)}$. The modifications when $p,q $ equal 1 or $\infty$ are obvious.

 Next, if we represent $(p,q)$ by the point of coordinates $(1/p,1/q)$, we may associate
all the spaces $L^{(p,q)} \; (1 \leqslant p,q \leqslant \infty)$ in a one-to-one fashion with the points of a
unit square $J = [0,1]\times [0,1]$  (see Figure 2). Thus, in this picture, the spaces  $L^p$ are on
the main diagonal, intersections $L^p \cap L^q$ above it and sums $L^p + L^q$ below. The space
$L^{(p,q)}$ is contained in $L^{(p',q')}$ if $(p,q)$ is on the left and/or above $(p',q')$. Thus the
smallest space $L^{(\infty,1)} = L^{\infty} \cap L^{1}$ corresponds to the upper  left
corner, the largest one, $L^{(1,\infty)} = L^{1} + L^{\infty}$, to the lower right corner. Inside the
square, duality corresponds to symmetry with respect to the center  
(1/2,1/2) of the square,
which represents the space $L^2$.

\setlength{\unitlength}{1cm}

\begin{picture}(17,12)
\put(3,1){\begin{picture}(17,12)

\put(0,0){\vector(0,1){9}}
\put(-0.2,9.5){\shortstack{$1/q$}} 

\put(0,0){\vector(1,0){10}}
\put(10.7,0){\makebox(0,0){$ 1/p$}}

\put(0,0){\line(1,1){8}}
\put(8,0){\line(0,1){8}}
\put(0,8){\line(1,0){8}}

\put(3,3){\dashbox{0.1}(3,3)}
\put(3,6){\line(1,-2){2}}

\put(3,0){\dashbox{0.1}(0,8)}
\put(0,6){\dashbox{0.1}(8,0)}

\put(-0.5,-0.4){\makebox(0,0){$ L^{\infty} $}}
\put(8.5,8.4){\makebox(0,0){$ L^{1} $}}

\put(8.1,-0.4){\makebox(0,0){$L^{(1,\infty)} = L^{1} + L^{\infty}$}}
\put(3.1,-0.4){\makebox(0,0){$L^{(p,\infty)} = L^{p} + L^{\infty}$}}
\put(-1.7,8.4){\makebox(0,0){$L^{(\infty,1)} = L^{\infty} \cap L^{1}$}}
\put(-1,6.2){\makebox(0,0){$ L^{\infty} \cap L^{q}$}}
\put(9.8,6.2){\makebox(0,0){$L^{(1,q)} = L^{1} + L^{q}$}}
\put(3,8.4){\makebox(0,0){$ L^{p} \cap L^{1}$}}

\put(4.1,4.5){\makebox(0,0){$L^2$}}
\put(6,6.4){\makebox(0,0){$ L^{q} $}}
\put(2.7,3.2){\makebox(0,0){$ L^{p} $}}
\put(3,6.4){\makebox(0,0){$L^p \wedge L^q = L^{(p,q)} $}}
\put(6.5,2.7){\makebox(0,0){$L^p \vee L^q = L^{(q,p)} $}}

\put(4,4){\makebox(0,0){$\scriptstyle\bullet$}}
\put(0,0){\makebox(0,0){$\scriptstyle\bullet$}}
\put(3,0){\makebox(0,0){$\scriptstyle\bullet$}}
\put(8,0){\makebox(0,0){$\scriptstyle\bullet$}}
\put(0,8){\makebox(0,0){$\scriptstyle\bullet$}}
\put(3,8){\makebox(0,0){$\scriptstyle\bullet$}}
\put(8,8){\makebox(0,0){$\scriptstyle\bullet$}}
\put(0,6){\makebox(0,0){$\scriptstyle\bullet$}}
\put(3,6){\makebox(0,0){$\scriptstyle\bullet$}}
\put(6,6){\makebox(0,0){$\scriptstyle\bullet$}}
\put(8,6){\makebox(0,0){$\scriptstyle\bullet$}}
\put(6,3){\makebox(0,0){$\scriptstyle\bullet$}}
\put(3,3){\makebox(0,0){$\scriptstyle\bullet$}}
\put(5,2){\makebox(0,0){$\scriptstyle\bullet$}}

\put(5.5,1.5){\makebox(0,0){$L^{\bar p} 
\vee L^{\bar q} = (L^p \wedge L^q)' $}}

\end{picture}}

\end{picture}

\begin{center}
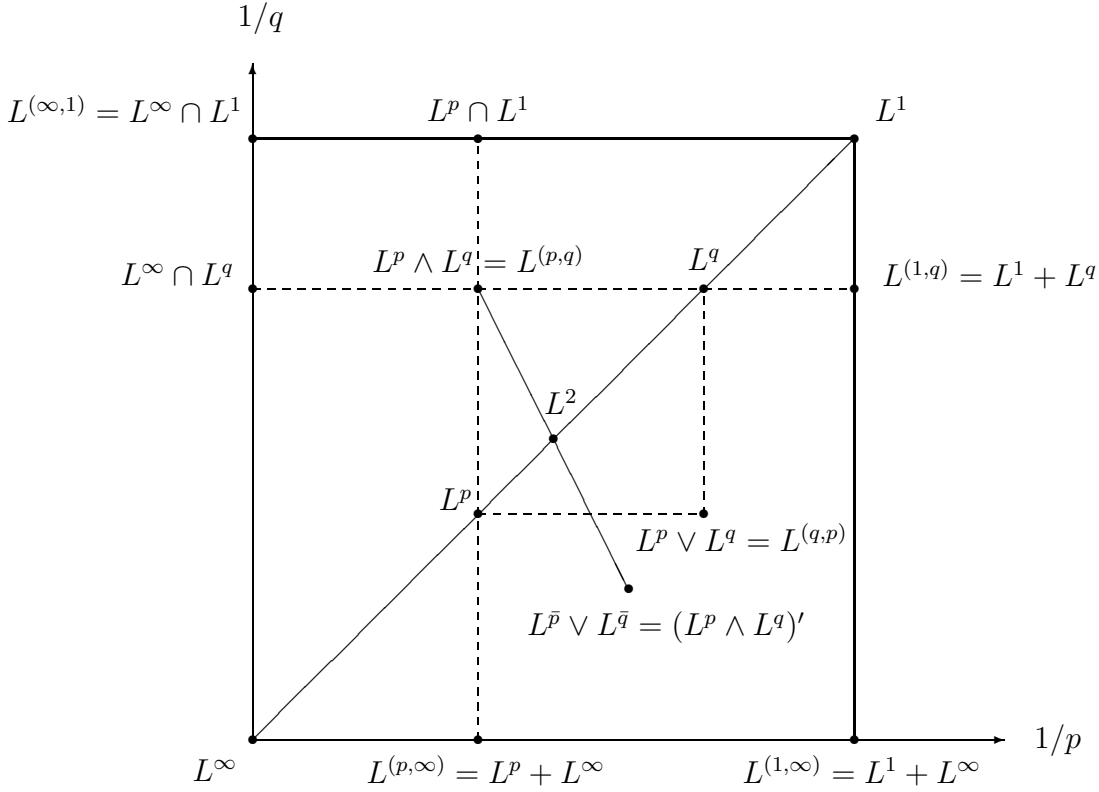
Figure 2 : The unit square describing the lattice $J$
\end{center}
\bigskip

\noindent
The ordering of the spaces corresponds to the following rule:
\be
L^{(p,q)} \subset L^{(p',q')} \quad \Longleftrightarrow \quad (p,q) \leqslant (p',q')
\quad \Longleftrightarrow \quad p \geqslant p' \; \mbox{ and } q \leqslant q'.
\label{order}
\en

For $\infty > q_o >1$, consider now the horizontal row $q = q_o, 
\, \{ L^{(p,q_o)} : \infty > p > 1 \}$. It corresponds to the chain:
\bea \label{hchain}
 \ldots \;\subset\; L^{r} \cap L^{q_o}
 \;\subset\;  \ldots \;\subset\;  L^{q_o} \;\subset\; \ldots \;\subset\; 
L^{s} + L^{q_o} \;\subset\; \ldots   
\\
\makebox[5cm]{} (\infty > r > q_o > s > 1)  \nonumber
\ena 
sitting between the extreme elements $L^{\infty} \cap L^{q_o}$ on the left and $ L^{1} + L^{q_o}$ 
on the right. The point is that all the embeddings in the chain (\ref{hchain}) are continuous and
have dense range.

 The same holds true for a vertical row $p = p_o, 
\, \{ L^{(p_o,q)} : 1 < q < \infty  \}$:
\bea \label{vchain}
 \ldots \;\subset\; L^{p_o} \cap L^{s}
 \;\subset\;  \ldots \;\subset\;  L^{p_o} \;\subset\; \ldots \;\subset\; 
L^{p_o} + L^{r} \;\subset\; \ldots   
\\
\makebox[5cm]{} ( 1 < s < p_o < r < \infty)  \nonumber
\ena 
Combining these two facts, we see that the partial order extends to the spaces 
$L^{(p,q)} \; (1 < p,q < \infty)$, inclusion meaning now continuous embedding with 
 dense range.

Now the set of points contained in the square $J$ may be considered as an involutive lattice with
respect to the partial order (\ref{order}), with operations:
\beano
(p,q) \wedge (p',q') &=& (p \vee p',q \wedge  q') \\
(p,q) \vee (p',q') &=& (p \wedge p',q \vee q') \\
\overline{(p,q)}  &=& (\bar p,\bar q),
\enano
 where, as usual,  $p \wedge p'=  \min \{p,p'\}, \; p \vee p' =  \max \{p,p'\}$.

The considerations made above imply that the lattice \J\ generated by $\I = \{L^p\}$ is already
obtained at the first generation. For example,
$L^{(r,s)} \wedge L^{(a,b)} = L^{(r\vee  a,s \wedge b)}$  (see Figure 3), 
and the latter may be either above, on or below the diagonal, depending on the values of the
indices. For instance, if $p < q < s$, then $L^{(p,q)} \wedge L^{(q,s)} = L^{q}$, both as sets and
as topological vector spaces.

The conclusion is that, using this language, the only difference between the two cases 
$\{L^p([0,1])\}$ and $\{L^p(\BR)\}$ lies in the type of order obtained: a chain $I$ (total order)
or a partially ordered lattice $J$. From this remark,
 the lattice completion of \J\ can be
obtained exactly as before, using the results of \cite{dav}. This introduces again Fr\'echet and
DF-spaces, all reflexive if we start from $1<p< \infty$, and in natural duality as in the previous
case. In particular, for the spaces of the first `generation', it suffices to consider intervals $S
\subset [1,\infty]$ and define the spaces
$$
L^P(S) = \bigcap_{q\in S}  L^q, \quad L^I(S) = \bigcup_{q\in S}  L^q. 
$$
Then:
\bei
\item
If $S$ is a closed interval $S=[p,q]$, with $p<q$, then  $L^P(S)=L^p \wedge L^q = L^{(q,p)}$ and
$L^I(S)=L^p
\vee L^q = L^{(p,q)}$ are Banach spaces. 
If $S$ is a semi-open or open interval, $L^P(S)$ is a non-normable Fr\'echet space and $L^I(S)$ 
a DF-space.
\item
Let $S \subset (1,\infty) $ and define ${\overline S} = \{ \bar q : q \in S\}$. Then
$(L^P(S))' =  L^I({\overline S}), \quad (L^I(S))' = L^P({\overline S})$
\eni
\setlength{\unitlength}{1cm}

\begin{picture}(17,12)
\put(3,1){\begin{picture}(17,12)

\put(0,0){\vector(0,1){9}}
\put(-0.2,9.5){\makebox(0,0){$1/q$}} 

\put(0,0){\vector(1,0){9}}
\put(9.7,0){\makebox(0,0){$ 1/p$}}

\put(0,0){\line(1,1){8}}
\put(8,0){\line(0,1){8}}
\put(0,8){\line(1,0){8}}

\put(0,2){\dashbox{0.1}(3.52,0)}
\put(0,5){\dashbox{0.1}(6,0)}
\put(3.5,2){\dashbox{0.1}(0,6)}
\put(6,5){\dashbox{0.1}(0,3)}

\put(-0.4,-0.4){\makebox(0,0){$ {\infty} $}}
\put(8,-0.4){\makebox(0,0){$ 1 $}}

\put(5.6,6){\makebox(0,0){$L^a$}}
\put(4.9,5.4){\makebox(0,0){$L^b$}}
\put(3.1,3.5){\makebox(0,0){$L^r$}}
\put(1.9,2.4){\makebox(0,0){$L^s$}}
\put(4,1.6){\makebox(0,0){$L^{(r,s)}=L^r + L^s $}}
\put(5.4,3.9){\shortstack{$L^{(a,b)}$\\ \quad$=L^a + L^b$}}

\put(3.5,8.3){\makebox(0,0){$r$}}
\put(6,8.3){\makebox(0,0){$a$}}
\put(3.5,-0.4){\makebox(0,0){$r$}}
\put(6,-0.4){\makebox(0,0){$a$}}
\put(3.5,0){\makebox(0,0){$\scriptstyle\bullet$}}
\put(6,0){\makebox(0,0){$\scriptstyle\bullet$}}
\put(-0.4,2){\makebox(0,0){$s$}}
\put(-0.4,5){\makebox(0,0){$b$}}
\put(-0.4,8){\makebox(0,0){1}}

\put(0.3,5.3){\shortstack{$L^{(r\vee a, s\wedge b)}= L^{(r,b)}$ \\ 
$\qquad =L^r \wedge L^b $}}

\put(5,5){\makebox(0,0){$\scriptstyle\bullet$}}
\put(0,0){\makebox(0,0){$\scriptstyle\bullet$}}
\put(6,6){\makebox(0,0){$\scriptstyle\bullet$}}
\put(3.5,5){\makebox(0,0){$\scriptstyle\bullet$}}
\put(6,5){\makebox(0,0){$\scriptstyle\bullet$}}
\put(3.5,3.5){\makebox(0,0){$\scriptstyle\bullet$}}
\put(2,2){\makebox(0,0){$\scriptstyle\bullet$}}
\put(3.5,2){\makebox(0,0){$\scriptstyle\bullet$}}

\end{picture}}

\end{picture}

\begin{center}
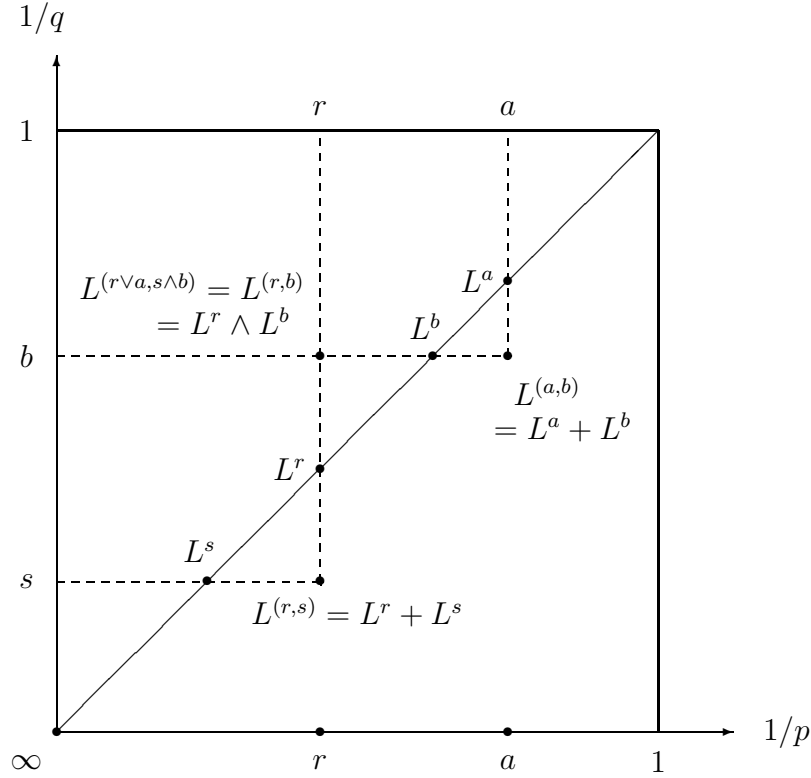
Figure 3 : The intersection of two spaces from $J$.
\end{center}
\bigskip

\noindent
A special r\^ole will be played in the sequel by the spaces $L^I$ corresponding to semi-infinite
intervals, namely:
\beano
L^{(p,\infty)} &=& L^I([p,\infty]) \;=\; \bigcup_{p\leqslant s \leqslant \infty}  L^s\;=\; L^p + L^\infty, \;
\mbox{which is a nonreflexive Banach space}. \\
L^{(p,\omega)} &=& L^I([p,\infty)) \;=\; \bigcup_{p\leqslant s < \infty}  L^s, \;
\mbox{which is a reflexive DF-space}.
\enano

As for the lattice completion \F, one can essentially repeat the argument of \cite[Example 3.B]{PIP3}
and build an `enriched' or `nonstandard' square $F$, exactly as in the previous section.
Take first $1<q<\infty$, that is, the interior $J_o$ of the square $J$. 
The extreme spaces of the corresponding complete lattice $\F_o$ are:
$$ 
V_o^\# = \bigcap_{1<q<\infty} L^q, \quad \mbox{ and } \quad V_o = \bigcup_{1<q<\infty} L^q = \sum_{1<q<\infty} L^q,
$$
with their projective and inductive topologies, respectively. All embeddings are
continuous and have dense range. Thus the space $V_o$, together with either of the two lattices $\J_o
=\{V_\alpha,\, \alpha \in J_o\}$ or 
$\F_o =\{V_\alpha, \, \alpha \in F_o \}$, is a {PIP}-space, with the usual $L^2$ inner
product and $(V_\alpha)^\# =  (V_\alpha)^\times = (V_{\bar\alpha})$.
 
Similar results 
are valid when one includes $L^1$ and $L^\infty$, except for the obvious modifications concerning
duality. The extreme spaces of the full lattice \F\ then are
$V_J^\# = L_\rho^\# = L^1 \cap L^\infty$ and 
$V_J = L_\rho = L^1 + L^\infty$, with their projective and inductive
norms,  which make them into nonreflexive Banach spaces (none of them is the dual of the other).
Notice that the space $L_\rho$, originally introduced by Gould \cite{zaanen}, contains strictly
all the $L^p, \; 1 \leqslant  p \leqslant \infty$.

We turn now to the \pa\ structure on $V_J$. Again we start from the lattice \J, which is generating
for the { PIP}-space structure. The basic fact  is H\"older's inequality, which says
that pointwise multiplication is continuous from $L^p \times L^q$ into $L^r$, where 
$1/p + 1/q = 1/r$. From this we can compute the multipliers of all the elements of \J\ in several
steps (as usual we write 
$p \wedge q=  \min \{p,q\}, \; p \vee q =  \max \{p,q\})$:
\bei
\item
$L^s \subset ML^p \;$ iff $\; \bar p \leqslant s \leqslant \infty.$  Thus   
$$
ML^p = \bigcup_{s\geqslant \bar p}L^s = L^{(\bar p,\infty)}.
$$

\item
Let $p>q$, so that $ L^{(p,q)} =  L^p \wedge  L^q.$ Then 
$$
ML^{(p,q)} = ML^p \vee ML^q =  L^{(\bar p,\infty)}\vee  L^{(\bar q,\infty)}= L^{(\bar p,\infty)}.
$$
\item
Let $p<q$, so that $ L^{(p,q)} =  L^p \vee  L^q.$ Then 
$$
ML^{(p,q)} = ML^p \wedge ML^q =  L^{(\bar p,\infty)}\wedge L^{(\bar q,\infty)}= L^{(\bar q,\infty)}.
$$
\item
Thus in all cases
\be
ML^{(p,q)} = L^{({\bar p}\wedge {\bar q},\infty)} =  L^{(\overline{p \vee q},\infty)}.
\label{multpq}
\en
\eni
If one does not want to include $L^{\infty}$, one simply replaces (\ref{multpq}) by
\be
ML^{(p,q)} = L^{({\bar p}\wedge {\bar q},\omega)} =  L^{(\overline{p \vee q},\omega)}.
\label{multomega}
\en
Applying the rule (\ref{multpq}) or(\ref{multomega})  twice, one gets immediately 
$MML^{(p,q)} = L^{(p\vee q,\infty)}$, resp. $MML^{(p,q)} = L^{(p\vee q,\omega)}$.

In conclusion, the generating family for multiplication is the set 
$\J^M = \{ L^{(p,\infty)}, \; 1 \leqslant p \leqslant \infty\}$, corresponding to the bottom side of the
square $J$ in Figure 2, and it is a chain of Banach spaces, exactly as in the case of the $L^p$
spaces over $[0,1]$. Thus we write the partial multiplication on $L^1 + L^\infty$ as:
\be
f \in M(g) \; \Leftrightarrow \; \exists \, q \in [1, \infty] \; \mbox{ such that }
f \in L^{(q,\infty)}, \, g \in L^{(\bar q,\infty)}, \;1/q + 1/{\bar q} = 1,
\label{multf}
\en
and that on $V$ as:
\be
f \in M(g) \; \Leftrightarrow \; \exists \, q \in (1, \infty) \; \mbox{ such that }
f \in L^{(q,\omega)}, \, g \in L^{(\bar q,\omega)}, \;1/q + 1/{\bar q} = 1,
\label{multV}
\en
Finally,  we can immediately conclude that the complete lattice
$\F^M$ is the `enriched' chain $\F^M = \{ L^{(p\epsilon,\infty)}, 
\; p\epsilon = p-, p \, \mbox{ or }\, p+, \; 1 \leqslant p \leqslant \infty\}$, and similarly with $\omega$
instead of $\infty$.

Exactly as in the case of a finite interval, we may restrict  the generating spaces to 
$\{L^s, \; p \leqslant s \leqslant \infty)$, which amounts to take a subsquare of $J$. The rest is obvious.

Another interesting structure of \pa\ may be given to the spaces $L_\rho$ or $V$, simply replacing
multiplication by convolution. According to  Hausdorff-Young's inequality, convolution maps
  $L^p \times L^q$ continuously into $L^r$, where 
$1/p + 1/q = 1 + 1/r$. From this we can compute the multipliers of all the elements of \J\ as in the
previous case (to avoid confusion, we use here the notation $M_*$):
\bei
\item
$L^s \subset M_*L^p \;$ iff $\;  s \leqslant \bar p.$  Thus   
$$
M_*L^p = \bigcup_{s\leqslant \bar p}L^s = L^{(1,\bar p)} = L^1 + L^{\bar p}.
$$

\item
For $p>q, \;   M_*L^{(p,q)} = L^{(1,\bar p)}$,
and for $p<q, \; M_*L^{(p,q)} = L^{(1,\bar q,)}$.
\item
Thus in all cases
\be
M_*L^{(p,q)} = L^{(1, {\bar p}\wedge {\bar q})} =  L^{(1,\overline{p \vee q})}.
\label{multpqconv}
\en
\eni
Again these multiplier spaces constitute a chain, this one  corresponding to the right-hand side of
the square $J$.

\subsection{\hspace{-4mm}. Amalgam spaces} 

The lesson of the previous example is that an involutive  lattice of (preferably reflexive)
Banach spaces (that is, a { PIP}-space of type B or H \cite{PIP4}) turns quite naturally into a 
(tight) \tpa\ if it possesses a partial multiplication that
verify a (generalized) H\"older inequality. A whole class of examples is given by the so-called
{\em amalgam spaces} first introduced by N. Wiener \cite{wiener} and developed systematically by
Holland \cite{holland}. The simplest ones are the spaces 
$(L^p,\ell^q)$ (sometimes denoted $W(L^p,\ell^q)$) consisting of functions on $\BR$ which are 
locally in $L^p$ and have 
$\ell^q $ behavior at infinity, in the sense that the  $L^p$ norms over the intervals 
$(n,n+1)$ form an $\ell^q $ sequence (see the review paper \cite{four}).
For $1 \leqslant p,q < \infty$, the norm
$$
\| f \|_{p,q} = \left\{ \sum_{n=-\infty}^{\infty}
     \left[\int_n^{n+1}|f(x)|^p dx\right]^{q/p}\right\}^{1/q}
$$ 
makes $(L^p,\ell^q)$ into a Banach space. The same is true for the obvious extensions to $p$ and/or
$q$ equal to 1 or $\infty$. Notice that $(L^p,\ell^p) = L^p$. The  spaces 
$(L^p,\ell^q)$ have many interesting applications, for instance in the context of various Tauberian theorems. 
New ones have been  found recently   in the theory of frames (nonorthogonal expansions) \cite{bened}.

These spaces obey the following (immediate) inclusion relations, with all embeddings continuous:
\bei
\item
If $q_1 \leqslant q_2$, then $(L^p,\ell^{q_1}) \; \subset \;(L^p,\ell^{q_2}).$ 
\item
If $p_1 \leqslant p_2$, then $(L^{p_2},\ell^{q}) \; \subset \;(L^{p_1},\ell^{q}).$
\eni 
From this it follows that the smallest space is $(L^\infty,\ell^1)$ and the largest one is
$(L^1,\ell^\infty)$, and therefore
\bei
\item
If $p\geqslant q$, then $(L^p,\ell^q) \; \subset\; L^p \cap L^q \; \subset\;L^s, \, \forall \, q<s<p$.
\item
If $p\leqslant q$, then $(L^p,\ell^q) \; \supset\; L^p \cup L^q $.
\eni
Once again, H\"older's inequality is  satisfied. Whenever $f \in (L^p,\ell^q)$ and
$g \in (L^{\bar p},\ell^{\bar q})$, then $fg\in L^1$ and one has
$$
\|f g\|_1 \leqslant \|f\|_{p,q} \, \|g\|_{\bar p,\bar q}.
$$
Therefore, one has the expected duality relation:
$$
(L^p,\ell^q)' = (L^{\bar p},\ell^{\bar q}), \; \mbox{ for } 1 \leqslant q,p < \infty.
$$
The interesting fact is that, for $1 \leqslant p,q \leqslant \infty$, the set \J\ of all amalgam spaces 
$\{(L^p,\ell^q)\}$ may be represented by the points $(p,q)$ of the {\em same} unit square $J$ as in
the previous example, with the {\em same} order structure. In particular, \J\ is a lattice with
respect to the order (\ref{order}):
\beano
(L^p,\ell^q) \wedge (L^{p'},\ell^{q'}) &=& (L^{p\vee p'},\ell^{q \wedge q'})
\\
(L^p,\ell^q) \vee (L^{p'},\ell^{q'})&=& (L^{p \wedge p'},\ell^{q \vee q'}),
\enano
where again $\wedge$ means intersection with projective norm and 
 $\vee$ means vector sum with inductive norm.

We turn now to the \pa\ structure of \J. At first sight, the situation becomes different, because
whereas $L^1$ is a \pa, $\ell^\infty$ is an algebra under componentwise multiplication, 
$(a_n)\cdot( b_n)= (a_n b_n)$.
The $L^p$ component characterizes the local behavior. Hence,
$$
M(L^p,\ell^q) \supset (L^{\bar p},\ell^{\infty}), \;  \forall \, q,  
$$
and since the latter are totally ordered, we obtain, exactly as in the cases of the $L^p$ spaces:
$$
M(L^p,\ell^q) = (L^{\bar p},\ell^{\infty}),  
$$
Thus the natural partial multiplication on \J\ reads:
\be
f \in M(g) \; \Leftrightarrow \; \exists \, p \in [1, \infty] \; \mbox{ such that }
f \in (L^{p},\ell^{\infty}) \; \mbox{ and } \; g \in (L^{\bar p},\ell^{\infty}). 
\label{multamal}
\en
The rest is as before, including the identification of the complete lattice \F\ with the
`enriched' interval $[1,\infty]$.

Since the amalgam spaces $ (L^p,\ell^q) $ obey the same Hausdorff-Young inequality as the 
$L^p$ spaces, we may obtain here too another structure of \pa\ with the convolution as partial
multiplication. Let $f \in (L^p,\ell^q)$ and $g \in (L^{p'},\ell^{q'})$, with
$1/p + 1/{p'} \geqslant 1, \;  1/q + 1/{q'} \geqslant 1$, that is, $p' \leqslant {\bar p}, \; q' \leqslant {\bar q}$.
Then $f * g \in (L^{p''},\ell^{q''})$, with $1/{p''} = 1/p + 1/{p'} - 1, \; 
1/{q''} = 1/q + 1/{q'} - 1.$ By the same arguments as in the previous section, we obtain
\be
M_*(L^p,\ell^q) = (L^1,\ell^{{\bar p}\wedge {\bar q}})
 = (L^1,\ell^{\overline{p \vee q}})
\en
As before, these multiplier spaces constitute a chain, corresponding to the right-hand side of the
square $J$.

\section{\hspace{-5mm}.  Examples II: Topological \pa s of operators} 

\subsection{\hspace{-4mm}. Operators on a lattice of Hilbert  spaces } 
 
Our first example is the \pa\ of operators on a lattice of Hilbert  spaces (LHS), also called 
indexed { PIP}-spaces of type (H) \cite{PIP4}. By this we mean a vector space $V$ together with a
 family of subspaces $V_I = \{\H_r, \, r \in I \}$, where
\bei
\item
$V = \sum_{r\in I}\H_{r}$;

\item
the index set $I$ is an involutive lattice with order-reversing involution $r \leftrightarrow \overline r$
(that is, $p \leqslant q $ implies 
$\overline q \leqslant \overline p $ and $ \overline {\overline p} = p$) and a unique element $o$ such that ${\overline o} = o$;

\item
each $\H_r$ is a Hilbert space with norm $\| \cdot \|_r$ and 
$\H_{\overline r} = \H_{r}^{\times}$,
the anti-dual of $\H_{r}$; in particular, 
$\H_{\overline o} = \H_{o}^{\times} = \H_{o}$ ;

\item
the family $V_I$ is an involutive lattice under set inclusion and lattice operations
\bei
\item[.]
$\H_{p\wedge q} = \H_{p} \cap \H_{q}$, with the projective norm
$\| f \|_{p\wedge q}^2 = \| f \|_{p}^2  + \| f \|_{q}^2, $

\item[.]
$\H_{p\vee q} = \H_{p} + \H_{q}$, with the inductive norm
$\| f \|_{p\vee q}^2 =  \inf_{f=g+h}\left(\| g \|_{p}^2  + \| h \|_{q}^2\right)$, \\ 
$(g \in \H_{p}, f \in \H_{q})$

\eni
(we  use squared norms in these definitions in order to get Hilbert norms for the projective and inductive ones).  

\item
The inner product  of $\H_{o}$ extends to a partial 
inner product $\langle \cdot | \cdot \rangle$, that is, a Hermitian sesquilinear form   defined exactly on dual pairs 
$\H_{r}, \H_{\overline r}$.
\eni
It follows that 
$(V, \langle \cdot |\cdot\rangle)$ is a {PIP}-space, and 
$ V^{\#} = \bigcap_{r\in I}\H_{r}$. We assume
 the partial inner product to be nondegenerate, that is, 
$(V^{\#})^{\bot} = \{0\}$, which means that $ \langle f \,| \, g \rangle = 0, \forall \, f \in 
V^{\#},$ implies $ g = 0.$ This  entails
that  $( V^{\#}, V)$, as well as  every pair 
 $(\H_{r}, \H_{\overline r})$, is a
dual pair  in the sense of topological vector space theory  \cite{schaefer}.
Note that the Mackey topology 
$\tau(\H_{r}, \H_{\overline r})$ on $\H_{r}$ coincides with the original norm topology.

Once again the topological and lattice structures coincide: 
$q < p $ implies $\H_{q} \subset \H_{p}$ and the embedding is continuous with dense range.
Similarly, $\H_{p\wedge q}$ and  $\H_{\overline p\vee \overline q}$ are dual to each other. Moreover, $ V^{\#}$ is dense in every $\H_{r},\, r \in I$.

Typical examples of LHS are:
\bei
\item[(i)]
{\sl Hilbert scales }

Many examples of Hilbert scales, discrete or continuous, appear in applications. For instance:
\bei
\item[.]
The scale built on the powers of a positive self-adjoint operator $H > 1$:
$\H_{n} = D(H^n)$, with the graph norm $\| f \|_n = \| H^n f \|, \, n \in \BN$, and $\H_{-n} =\H_{n}^{\times}$.

\item[.]
The scale of Sobolev spaces $W^2_s(\BR), \, s \in \BR$, where $f \in W^2_s(\BR)$ if its Fourier transform $\widehat f$ satisfies the condition 
$(1 + |.|^2)^{s/2}\, \widehat f \in L^2(\BR)$. The norm is 
$\| f \|_s = \|  (1 + |.|^2)^{s/2}\, \widehat f \|, \, s \in \BR$. Of course, similar considerations hold for the Banach scale
$\{ W^p_s(\BR), \, s \in \BR\}, \, 1 < p < \infty$, but here we restrict ourselves to the Hilbert case $p=2$. 

\eni
We will come back to these two examples at the end of this Section 5.

\item[(ii)]
{\sl Weighted $\ell^2$ sequence spaces }

Given a sequence of positive numbers, $r = (r_n), \, r_n > 0$,  define
$\ell^2(r) = \{ x = (x_n) : \sum_{n=1}^\infty \ |x_n|^2 \,r_n^{-1} < \infty\}.$
The lattice operations read:
\bei
\item[.]
involution: $\ell^2(\overline{r}) = \ell^2(r)^\times, \, \overline{r}_n = 1/r_n$.
\item[.]
infimum: $\ell^2(p) \wedge \ell^2(q) = \ell^2(r), \, r_n = \min(p_n,q_n)$.
\item[.]
supremum: $\ell^2(p) \vee \ell^2(q) = \ell^2(s), \, s_n = \max(p_n,q_n)$.
\eni
As for the extreme spaces, it is easy to see that the family $\{\ell^2(r)\}$ generates the space 
$\omega$ of {\em all} complex sequences, while the intersection is the space $\varphi$ of all 
{\em finite} sequences.

\item[(iii)]
{\sl Weighted $L^2$ function spaces }

 Instead of sequences, we consider locally
integrable (i.e. integrable on bounded sets) functions $f \in L^1_{\rm loc}(\BR, dx)$ and define
again weighted spaces:
\beano
    I &=& \{ r  \in L^1_{\rm loc}(\BR, dx) : r(x) > 0, \; \mbox{a.e.} \}   \\
L^2(r) &=& \{ f  \in L^1_{\rm loc}(\BR, dx) : \int |f(x)|^2 \,r(x)^{-1}\, dx < \infty \}, 
\; r \in I.
\enano
Then we get exactly the same structure as in (ii):
\bei
\item[.]
involution: $L^2(r) \Leftrightarrow L^2(\overline{r}),   \; \overline{r} = 1/r$.
\item[.]
infimum: $L^2(p) \wedge L^2(q) = L^2(r), \, r(x) = \min(p(x),q(x))$.
\item[.]
supremum: $L^2(p) \vee L^2(q) = L^2(s), \, s(x) = \max(p(x),q(x))$.
\item[.]
extreme spaces: 
$$
\bigcup_{r\in I}L^2(r) = L^1_{\rm loc}, \quad \bigcap_{r\in I}L^2(r) = L^\infty_{\rm c},
$$
where $L^\infty_{\rm c}$ is the space of (essentially) bounded functions of compact support. The
central space is, of course, $L^2$.
\eni
An interesting subspace of the preceeding space is the LHS $V_\gamma$ generated by the weight
functions $r_\alpha (x) = \exp \alpha x$, for $-\gamma \leqslant \alpha \leqslant \gamma \, (\gamma>0)$.
 Then all the spaces of the lattice may be obtained by interpolation from 
$L^2(r_{\pm \gamma})$, and moreover,  the extreme spaces are themselves Hilbert spaces, namely  
\beano
V_\gamma^\# &=&  L^2(\BR, e^{-\gamma x} dx) \cap  L^2(\BR, e^{\gamma x} dx)
\;\; \simeq \;\; L^2(\BR, e^{-\gamma |x|} dx) \\
V_\gamma &=&  L^2(\BR, e^{-\gamma x} dx) +  L^2(\BR, e^{\gamma x} dx)
\;\; \simeq \;\;  L^2(\BR, e^{\gamma |x|} dx). 
\enano
This LHS plays an interesting role in scattering theory \cite{ant-goslar}.
\eni

\noindent
Actually the whole construction goes through if one takes for $\H_r$ a reflexive Banach space, as
in interpolation theory \cite{berghlof}. In this way one recovers the families $\{\ell^p\}$ or
$\{L^p\} \, (1<p<\infty)$ discussed in Section 4. For simplicity we restrict the discussion to a
LHS.

Let $V_I = \{\H_r, \, r \in I \}$ be a LHS. The whole idea behind this structure (as for general
{ PIP}-spaces) is that vectors should not be considered individually, but only in
terms of the subspaces $\H_r$, which are the building blocks of the theory.
The same spirit determines the definition of an operator on a LHS space: only bounded operators
between Hilbert spaces  are allowed, but an operator is a (maximal) coherent collection of these. To
be more specific, an {\em operator} on $V_I$ is a
map $A: \D(A) \to V$, such that:
\bei
\item[(i)]
$\D(A) = \bigcup_{q \in D(A)} \H_q$, where $D(A)$ is a nonempty subset of $I$.
\item[(ii)]
For every $q \in D(A)$, there is $p \in I$ such that the restriction  $A: \H_q \to \H_p$ is linear
and bounded (we denote it by $A_{pq} \in \B(\H_q,\H_p)$).
\item[(iii)]
$A$ has no proper extension satisfying (i) and (ii).
\eni
The bounded linear  operator $A_{pq}: \H_q \to \H_p$ is called a {\em representative} of
$A$. Thus $A$ is characterized by two subsets of $I$:\footnote{The set $I(A)$ was denoted $R(A)$ in \cite{PIP4}, but this obviously conflicts
with the space of right multipliers, to be defined below.}
\beano
D(A) &=& \{ q \in I : \;\mbox{there is a } \,   p \; \mbox{such that}\; A_{pq} \;\mbox{exists} \} 
\\ I(A) &=& \{ p \in I :\;  \mbox{there is a } \, q \; \mbox{such that}\; A_{pq} \;\mbox{exists}
\}  
\enano

\setlength{\unitlength}{1cm}

\begin{picture}(17,10)
\put(3,0.5){\begin{picture}(17,10)

\put(8,0){\vector(0,1){8}}
\put(8.03,2){\line(0,1){5.9}}
\put(8.4,8){\makebox(0,0){$p$}} 
\put(8,-0.4){\makebox(0,0){$I$}}

\put(0,1){\vector(1,0){10}}
\put(0,1.03){\line(1,0){6}}
\put(6,1.03){\makebox(0,0){$\shortmid$}}
\put(10.5,1){\makebox(0,0){$q$}}
\put(-0.5,1){\makebox(0,0){$I$}}

\put(0,2){\dashbox{0.1}(8.03,0)}
\put(0,4){\dashbox{0.05}(4.2,0)}
\put(4.2,4){\dashbox{0.05}(0,4)}
\put(6,1){\dashbox{0.1}(0,7)}

\bezier{800}(0,2.2)(5,2)(5.5,8)

\put(2,6){\makebox(0,0){$J(A)$}}
\put(-0.5,4.5){\makebox(0,0){$q'<q$}}
\put(3.5,8){\makebox(0,0){$p'>p$}}
\put(4.8,3.8){\makebox(0,0){$(q,p)$}}
\put(2.5,0.5){\makebox(0,0){$D(A)$}}
\put(6,0.7){\makebox(0,0){$q_{\rm max}$}}
\put(8.6,2){\makebox(0,0){$p_{\rm min}$}}
\put(8.6,5){\makebox(0,0){$I(A)$}}

\end{picture}}
\end{picture}

\begin{center}
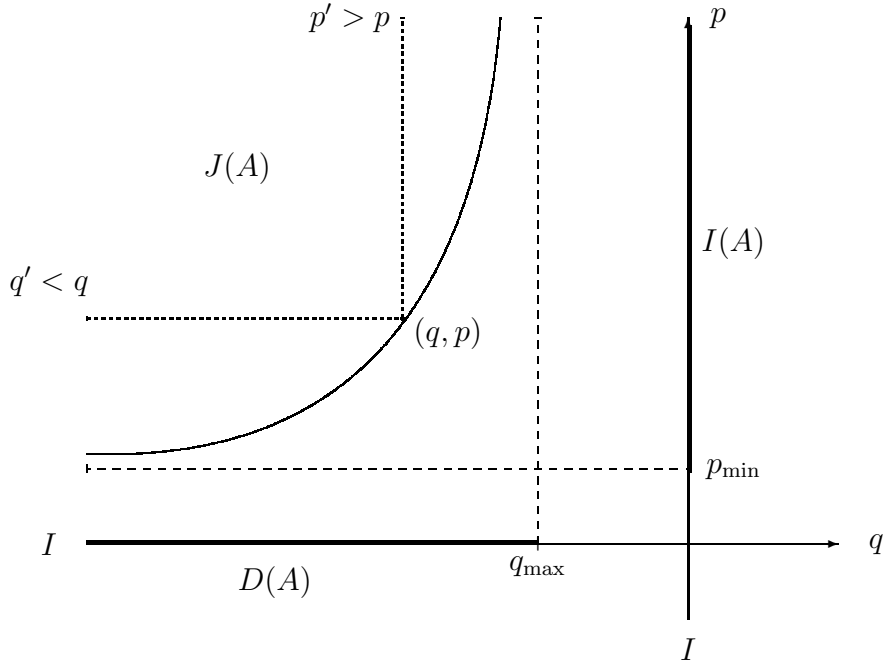
Figure 4 : The various sets characterizing the operator $A$ (in the case of a scale).
\end{center}
\bigskip

\noindent
   We denote by $J(A)$ the set of all such pairs
$(q,p)$ for which $ A_{pq}$ exists.
Thus the operator $A$ is equivalent to the collection of its
representatives
\be
A \simeq \{ A_{pq} : (q,p) \in J(A) \}.
\label{repres}
\en
$D(A)$ is an initial subset of $I$:
 if $q \in D(A)$ and $q' < q$, then $q' \in D(A)$, and $A_{pq'} = A_{pq}E_{qq'}$, where 
$E_{qq'}$ is the unit operator (this is what we mean by `coherent').
In the same way,  $I(A)$ is a final subset of $I$: if $p \in I(A)$ and $p' > p$, then $p' \in I(A)$. 
Figure 4 illustrates the situation in the case of a Hilbert scale ($I$ totally ordered).
Notice that, even then,  the extreme elements $q_{\rm max}=\bigvee_{q\in D(A)}q$, resp.
$p_{\rm min}=\bigwedge_{p\in I(A)}q$ need not belong to $D(A)$, resp.
$I(A)$, since $I$ is not a complete lattice in general. 
Also $J(A) \subset D(A) \times I(A)$, with strict inclusion in general.

 We denote by $Op(V_I)$ the set
of all operators on $V_I$. Since $V^\# \subset\H_r, \; \forall \, r \in I$, an operator may be
identified with a  sesquilinear form on $V^\# \times V^\#$.
Indeed, the restriction of any representative $A_{pq}$ to $V^\# \times V^\#$ is such a form, and they all coincide. 
Equivalently, an operator may be identified
with a linear map from $V^\#$ into  $V$. But the idea behind the notion of operator is to
keep also the algebraic operations on operators, namely:
\bei 
\item[(i)]{\sl  Adjoint $A\ha$ :}
every operator $A \in Op(V_I)$ has a unique adjoint $A\ha \in Op(V_I)$, defined by:
$$
\langle A\ha x \, | \, y \rangle = \langle  x \, | \, Ay \rangle , \;{\rm for}\,  y \in \H_r, \, r
\in J(A) \;{\rm and }\;\, x \in V_{\overline{s}}, \, s \in I(A),
$$
that is,    $(A\ha)_{\overline{r}\overline{s}} = (A_{sr})\ha \;$   (usual Hilbert space adjoint). 
This implies that 
$A\ha{}\ha = A, \, \forall \, A \in Op(V_I)$: no extension is allowed, because of the maximality
condition (iii).

\item[(ii)] {\sl Partial multiplication :}
$AB$ is defined iff there is a $q \in I(B) \cap D(A)$, that is, iff there is continuous 
factorization through some $\H_q$:
$$
\H_r \; \stackrel{B}{\rightarrow} \; \H_q \; \stackrel{A}{\rightarrow} \; \H_s , \quad
\mbox{i.e.} \quad  (AB)_{sr} = A_{sq} B_{qr}.
$$
\eni
Notice that here, contrary to the case of a general { PIP}-space, the domain $\D(A)$ is automatically
 a vector subspace of $V$. Therefore $Op(V_I)$ is a partial *-algebra (which means,
in particular, that the usual rule of distributivity is valid).

Now we turn to the spaces of multipliers. Our building blocks are the sets:
\be
\O_{pq} = \{ A \in Op(V_I) : A_{pq} \; \mbox{exists} \}.
\en
Clearly we have:
\bei
\item[$\scriptstyle\bullet$]
$L\O_{pq} \equiv L_p = \{C\in Op(V_I) : p \in D(C) \} 
=  \bigcup_s \B(\H_p, \H_s) \simeq {\rm End}\,(\H_p,V)$

\item[$\scriptstyle\bullet$]
$R\O_{pq} \equiv R_q = \{B\in Op(V_I) : q \in I(B) \} 
=  \bigcup_t \B(\H_t, \H_q) \simeq {\rm End}\,(V^\#,\H_q)$

\item[$\scriptstyle\bullet$]
$RL\O_{pq} = RL_p = R_p \in \F^R$

\item[$\scriptstyle\bullet$]
$LR\O_{pq} = LR_q = L_p \in \F^L.$

\eni
(in these relations, ${\rm End}\, (X,Y)$ denotes the space of all linear maps from $X$ into $Y$).

From this we deduce immediately, using the fact that $L,R$ are lattice anti-isomorphisms:
$$
\begin{array}{lllllll}
 L_p \wedge  L_q    &= &  L_{p\vee q} & \qquad   &L_p \vee  L_q    &= &  L_{p\wedge q} 
\medskip\\
 R_p \wedge  R_q    &= &  R_{p\wedge q} & \qquad  & R_p \vee  R_q    &= &  R_{p\vee q}. 
\end{array}
$$
In particular, $q \leqslant q'$ implies $R_q \subset R_{q'}$ and
$L_q \supset L_{q'}$.
Thus $\I^L = \{L_p\}$ is a sublattice of $\F^L$, $\I^R = \{R_p\}$ is a sublattice of $\F^R$, and
both are generating (except that they do not contain the extreme elements in general, see below). 
In addition $\I^L,\I^R$ consist of matching pairs $(R_q, L_q)$.
Indeed,  given $A \in Op(V_I)$, we may rewrite
\be
D(A) = \{q \in I | A \in L_q \}, \quad I(A) = \{p \in I | A \in R_p \}.
\en
and therefore
\be
A \in L(B) \quad \Leftrightarrow \quad \exists \, p \in I \; \mbox{ such that } 
A \in L_p, B \in R_p.
\label{compat}
\en
From (\ref{compat}), we deduce  individual multiplier spaces:
\be
L(A) = \bigvee _{p\in I(A)} L_p = L_{p_{\rm min}}\ , 
\quad R(A) = \bigvee _{q\in D(A)} R_q = R_{q_{\rm max}}\ .
\en
Note that these two subsets do not belong to   $\I^L$, resp. $\I^R$, in general, but to the complete
lattice generated by the latter.

In the same way, we obtain
\be
ROp(V_I) = \{ A | I(A) = I\} =  \bigcap_{r \in I} R_r,
\en
which may be identified with the space ${\rm End}\,(V^\#)$ of all linear maps from $V^\#$ into itself. Again, 
$ROp(V_I) \not\in \I^R$.
Similarly,
\be
LOp(V_I) =  \{ A | D(A) = I\}   
= \bigvee_{p \in I} L_p \simeq {\rm End}\,(V).
\en

The final point concerns topologies on spaces of multipliers. As a consequence of the
identification (\ref{repres}) of an operator with the set of its representatives, the
\pa\ $Op(V_I)$ itself has the structure of an inductive limit of Banach spaces:
\be
Op(V_I) \simeq \bigcup_{q,p \in I} \B(\H_q, \H_p).
\label{op1}
\en
One may also consider the extreme spaces $ V^{\#} = \bigcap_{r\in I}\H_{r}$, 
$V = \sum_{r\in I}\H_{r}$. On $V$, the inductive limit topology coincides with  the Mackey 
topology $\tau (V, V^{\#})$, but on $ V^{\#}$, the projective topology 
may be coarser than the Mackey 
topology $\tau (V^{\#}, V)$. This gives another possibility of giving a topology to  $Op(V_I)$,
by identifying an operator with a continuous linear map from $V^{\#}$ into $V$
(each of them endowed with its own Mackey topology), that is: 
\be
Op(V_I) \simeq \L(V^{\#}, V).
\label{op2}
\en
These various possibilities may be different in  general, which makes the problem quite involved. Instead we will consider several simpler cases.

\bigskip


(1) First, suppose that the extreme spaces $ V^{\#}$ and $V$ are themselves Hilbert spaces, as
for the LHS  $V_\gamma$ described above, or, in the Banach case, the lattices 
$\{\ell^p \}, \, \{L^p[0,1] \},$ $\{L^p(\BR) \}, \,\{ (L^p, \ell^q) \}$. In that case, 
the relation (\ref{op2}) gives immediately the identification $Op(V_I) \simeq \B(V^{\#}, V)$,
with its usual norm topology.
Similarly, $ROp(V_I) \simeq \B(V^{\#})$ and $LOp(V_I) \simeq \B(V)$.
More generally
\beano
L_q &=& \mbox{ind} \lim_t \B(\H_q, \H_t)\;\;\simeq \;\;\B(\H_q, V), \\
R_q &=& \mbox{ind} \lim_s \B(\H_s, \H_q)\;\;\simeq \;\;\B(V^{\#}, \H_q),
\enano
and these norm topologies coincide with the topologies $\lambda$, resp.
$\rho$, on $L_q$, resp. $R_q$. Finally the involution is clearly continuous on $Op(V_I)$, so that $Op(V_I)$ is a \tpa. However, tightness is open in general.

\bigskip

(2) The situation is still simple, and most of the results of (1) survive,  when $V_I$  consists of a scale (either continuous or discrete)
of Hilbert spaces. Then, indeed,  $I$ contains a countable subset $J$, stable  under the involution, and coinitial to $I$,  which means that,
 for each $r \in I$, there exists $q \in J$ such that $q \leqslant r$
($J$ is then automatically cofinal to $I$ : $\forall \, r \in I$, there exists $p \in J$ such that $r \leqslant p$). As a consequence,
 the projective topology $t_I$, defined by
$V_I$, is equivalent to that defined by $V_J=\{ \H_s, s \in J\}$. 
In this case 
\be 
V^{\#} = \bigcap_{s \in J} \H_s \ , 
\hspace{15 mm} V= \bigcup_{s \in J} \H_s \ ,  
\en 
and hence $V^{\#}$ is a reflexive Fr\'echet space and $V$ is a reflexive  DF-space. 
Thus the projective topology on $V^{\#}$ coincides with the 
Mackey  topology $\tau (V^{\#}, V)$, and no pathology arises. The
space $\L(V^{\#}, V)$ of Mackey continuous operators coincides exactly  with the
space of all linear maps from $V^{\#}$  into $V$ which are continuous from
$V^{\#}[t_I]$ into $V[t'_I]$, where $ [t'_I]$ denotes the strong dual topology.
In this situation, the space $\L(V^{\#}, V)$ provides an example of a
quasi*-algebra of operators \cite{lass3,ctrev} and the usual
theory applies.

In particular, in addition to (\ref{op2}), we have the identifications:
\be
ROp(V_I) \simeq \L(V^{\#}), \quad LOp(V_I) \simeq \L(V),
\label{Rop}
\en
where $\L(V^{\#})$ is the space of all continuous operators from
$V^{\#}[t_I]$ 
into itself and $\L(V)$ and the space of all continuous operators from
$V[t'_I]$
into itself (both these spaces can be identified with
subspaces of $\L(V^{\#}, V)$). Similarly one gets:
\be
L_q \simeq \L(\H_q, V), \quad
R_q \simeq \L(V^{\#}, \H_q), \quad q \in I.
\label{multscale}
\en
Of course, these results remain valid if  $I$ is not a scale, but a
lattice containing a countable subset $J = \overline{J}$, coinitial to $I$ : 
$V^{\#}$ is  Fr\'echet and $V$  a DF-space.

Topologies on  $\L(V^{\#}, V)$ can then be introduced following
\cite{lass3}. The most interesting seems to be the {\em uniform}
topology defined by the set of seminorms
$$ 
A \in \L(V^{\#}, V) \mapsto 
\sup_{f,g \in {\cal M}} |\langle f | A g \rangle | 
$$
where $ {\cal M}$ is a bounded subset of $V^{\#}[t_I]$. 
Then there are several possible ways of turning 
 $\L(V^{\#}, V)$ into a partial *-algebra, in such a way that one always has, as in (\ref{Rop}):
\be
 R\L(V^{\#}, V) = \L(V^{\#}), \quad  L\L(V^{\#}, V) = \L(V) . 
\label{LL}
\en
Since the involution and the multiplications are continuous with
respect to the uniform topology, $\L(V^{\#}, V)$ becomes a
topological \pa, no matter {\em how many} Hilbert spaces we use to
define (by composition) the multiplication (provided that the relations 
(\ref{LL}) are satisfied).
The simplest possibility,  usually adopted in the theory of
quasi *-algebras,  consists in considering {\em none} of them: this choice
yields very poor lattices of multipliers (for instance $\J^R$
contains only $R\L(V^{\#}, V)$ and $\L(V^{\#}, V)$ itself).
With this {\em trivial} lattice of multipliers, $\L(V^{\#}, V)$ is
a tight topological \pa\ for well-behaved spaces $V^{\#}$, 
typically a Fr\'echet space whose topology is the projective topology generated by an O*-algebra. In that case indeed, both $\L(V^{\#})$ and $\L(V)$ are uniformly dense in $\L(V^{\#}, V)$ \cite{lass2,schm}.
 
But this 
was clearly not what we had in mind when we considered a LHS! We
were, in fact interested in finding a larger (and possibly the
largest) lattice of multipliers, making use of the factorization via
the spaces $\{\H_p, p\in I\}$ (this corresponds, of course, to the
possiblility of getting the largest possible set of multiplicable
pairs). As said before, in all these cases, $\L(V^{\#}, V)$ is a
 \tpa, but tightness is still to be proven. 
It is interesting to notice the analogy of this procedure of `enrichment' of the lattice of
multiplier spaces with the similar operation of refinement or coarsening of a compatibility
relation, which also leads to the construction of suitable lattices of subspaces, either
containing, or contained in, the corresponding lattice as a sublattice 
(see \cite{PIP3,ak=refin} for a systematic discussion). One should also beware of possible
pathologies linked to associativity, discovered by K\"ursten \cite{kursten}. 
 
 One of the most interesting cases for applications is that of the Hilbert scale built on the powers of a  self-adjoint operator $H>1$. That is, 
$ \I = \{ \H_s, \, s \in I \equiv \BR \;{\rm or} \; \BZ\}$, where $\H_s= D(H^s), \; s \geqslant 0$, with the graph norm, and $H_{-s} = \H_s^\times,
\, V^{\#} = \bigcap_{s\in I} \H_{s} = D^\infty(H), V = \sum_{s\in I} \H_{s}$. 
The partial multiplication in  $Op(V_I) \simeq \L(V^{\#}, V)$
is defined by continuous factorization through some $\H_s$:
$ A \cdot B $ is defined whenever there exists $s \in I$ such that 
$B \in \L(V^{\#},\H_s)$ and $A \in \L(\H_s,V)$. The spaces of multipliers themselves, given in (\ref{multscale}), form  scales: 
\be
\I^L = \{ L_s = \L(\H_s, V), \; s \in I \}, \quad
\I^R = \{R_s =\L(V^{\#}, \H_s), \; s \in I\}.
\en
In the case of a discrete scale, $I = \BZ$, the lattices $\I^L,\I^R$ are already complete. For instance, if $K $ is a subset of $\BZ$, bounded from above,  then $\bigcap_{n \in K} R_n = R_{n_K}$, with $n_K = \max K$.
For a continuous scale,  $I = \BR$, this is no longer the case, but the lattice completion is obtained exactly as in the case of the $L^p$ spaces described in Section 4, by `enriching' the line $\BR$. For instance,
$$
\H_{s-} = \bigcap_{r<s} \H_{r}, \quad \H_{s+} = \bigcup_{t>s}\H_{t}.
$$
With their  projective, resp. inductive topology, $\H_{s-}$ is a reflexive Fr\'echet space and $\H_{s+}$ is a reflexive DF-space. The rest is as before, duality relations and lattice completions. The interesting point is the following result.
\beprop
\label{prop-tight_tpa}
--
Let $\I = \{ \H_s, \, s \in I \equiv \BR \;{\rm or} \; \BZ\}$
be the Hilbert scale built on the powers of a  self-adjoint operator $H>1$, with $V^{\#} = D^\infty(H), \, V = \sum_{s\in I} \H_{s}$. 
Then, with  partial multiplication  defined by continuous factorization through the spaces $\H_s, \;Op(V_I) \simeq \L(V^{\#}, V)$ is a tight \tpa\ with respect to the uniform topology.
\enprop
The proof is given  in the Appendix. Here instead, let us consider the two examples already mentioned:
\bei
\item[(i)]
The Hilbert scale around $L^2(\BR,dx)$ built on the powers of the self-adjoint operator $H = \frac{1}{2}(-\frac{d^2}{dx^2} + x^2)$
(this is the Hamiltonian of a quantum mechanical harmonic oscillator in one dimension). Going to the limits $n \to \pm \infty$ yields 
$$
 V^{\#} = \bigcap_{n\in \BZ}\H_{n} = \S(\BR) \quad {\rm and} \quad 
V = \sum_{n\in \BZ}\H_{n} = \S'(\BR),
$$
Schwartz' spaces of smooth fast decreasing functions and tempered distributions, respectively. In fact, this scale may be used for a simpler formulation of the theory of tempered distributions, called the Hermite representation \cite{simon}. This example illustrate the usefulness of considering $Op(V_I) \simeq \L(\S,\S')$ as a \pa. 
As a consequence of Proposition \ref{prop-tight_tpa},
the space $ROp(V_I)= \L(\S)$  is dense in every $R_n= \L(\S,\H_n)$. Thus we have a tight \tpa, already studied in \cite{russo}.

\item[(ii)]
The Sobolev  scale $\{W^2_s(\BR^n), \, s \in \BR\}$ is also of this type, with $H = 1 - \Delta$, acting in $L^2(\BR^n,d^n x) \,(\Delta$ is the $n$-dimensional Laplacian). 
The operators on this scale are the building blocks of the theory of partial differential operators. Again the point of view of a \tpa\ might be useful in applications, in particular the tightness condition may offer useful approximations. Notice that, if we take together the scale  $\{W^2_s\}$ and its Fourier transform  $\{\widehat{W^2_s}\}$, we recover the Schwartz spaces 
$\S,\S'$ as extreme spaces.
\eni

In the general case, where $I$ does not contain a countable coinitial subset (or sublattice) $J$, things get quite involved. 
Standard examples are the full LHS of weighted $\ell^2$ or $L^2$ spaces described above. It is probably pointless to treat the problem in such generality.


\subsection{\hspace{-4mm}. Partial O\xas} 

        Let $\H$ be a complex Hilbert space with inner product $\langle \cdot | \cdot \rangle$ 
and $\D$ a dense subspace of $\H$.
 We denote by $ \L\ad(\D,\H) $
the set of all (closable) linear
operators $X$ such that $ {\D}(X) = {\D},\; 
{\D}(X\x) \supseteq {\D}.$ The set $ \L\ad(\D,\H ) $ is a  \pa\ 
 with respect to the following operations : the usual sum $X_1 + X_2 $,
the scalar multiplication $\lambda X$,
the involution $ X \mapsto X\ad = X\x \up {\D}$ and the {\em (weak)} partial
 multiplication
$X_1 \mult X_2 = {X_1}\ad\x X_2$, defined whenever $X_2$ is a weak right
 multiplier of $X_1, \, X_2 \in R^{\rm w}(X_1)$, that is, iff $ X_2 {\D}
\subset {\D}({X_1}\ad\x)$ and  $ X_1\x {\D} \subset {\D}(X_2\x).$ 
It is easy to check that $X_1 \mult X_2$ is well-defined iff there exists $C \in
\L\ad(\D,\H) $ such that 
\be 
\langle X_2 f | {X_1}\ad g\rangle = \langle Cf|g \rangle, \quad \forall f,g \in \D;
\label{alt}
\en 
in this case $X_1 \mult X_2=C$. When we regard $
\L\ad(\D,\H) $ as a  \pa\  with those operations, we denote it
 by $\LDH $. 

A \po\  on \D\ is a *-subalgebra \M\ of $\LDH $,
 that is, \M\ is a subspace of $\LDH $, containing the identity
and such that $X\ad \in \M\ $ whenever $X \in \M\ $ and $X_1 \mult X_2 \in \M$
for any $X_1, X_2 \in \M$ such that $X_2 \in R^{\rm w} (X_1).$ As for
$\LDH$ itself, it is the largest \po\  on the domain \D.
The sets $R \LDH$ of the universal right multipliers of $\LDH$ and 
$L \LDH$ of the universal 
left multipliers of $\LDH$ can be described as follows \cite{ak}:
$$
R \LDH  =\left\{B \in \L\ad(\D,\H): \overline{B} \in \B(\H), \; B\D \subset \D^{*}\right\},
$$
where
$$ 
\D^{*}= \bigcap_{A\in \L\ad(\D,\H)}D(A^*)
$$
and 
$$
L \LDH = \left\{B^* : B \in R \LDH\right\}
$$
(If $\D = \D^{*}$, $\LDH$ is said to be self-adjoint).

In order to introduce a topology on $\LDH$ it is convenient to endow \D\ with a topology which
makes each $A \in \LDH$ continuous.  This can be done 
by defining the topology on \D\ by the following family of seminorms
$$
f \mapsto \|Af\|, \quad A \in \LDH.
$$ 
This topology will be denoted in what follows by $t_{\L\ad}$. Clearly, $t_{\L\ad}$ is 
 the projective topology 
defined on \D\ by $\L\ad(\D,\H)$ and for this reason each $A \in \LDH$ is continuous 
from \D\ into \H.

We will now define three topologies on $\LDH$ and check whether $\LDH$ is a \tpa\ with respect to
them.

\noindent \underline{\it Quasi-uniform topology, $\tau_{*}$}

\noindent It is defined by the set of seminorms
$$ 
A \mapsto \sup_{f \in {\cal N}} (\|Af\|+\|A\ad f\|), \quad {\cal N} \mbox{ bounded in } \D[t_{\L\ad}].
$$
By the definition itself of $\tau_{*}$  it follows that the map $A \mapsto A\ad$ is continuous.

\noindent If $\M \in {\cal F}^R$, then the corresponding topology $\rho^*_{\m}$ is defined by the set of seminorms
$$
 A\in \M \mapsto \sup_{f \in {\cal N}} (\|(X\mult A)f\|+\|(A\ad \mult X\ad)f\|), 
\quad  X \in L\M, \,{\cal N} \mbox{ bounded in } \D[t_{\L\ad}].
$$

The following lemma, proved in \cite{ak,am}, shows that if $\LDH$ is self-adjoint, the first
two conditions of  Definition \ref{def3.5} are fulfilled if $\LDH$ is endowed with
$\tau_{*}$.


\belem
-- 
If $\LDH$ is self-adjoint, then the maps $A \mapsto X\mult A$ and $A \mapsto A\mult Y$ are 
$\tau_{*}$-continuous for all $X \in L \LDH$ and $Y \in R \LDH$; thus
$\LDH [\tau_{*}]$ is a  \tpa.    
\label{5.2}
\enlem
The topology on $\M\in {\cal F}^R$, defined as in Section 3, will be called here $\rho^*_{\m}$ to
remind its dependence on $\tau_{*}$. 
 Analogously, if $\N\in {\cal F}^L$, its topology will be called $\lambda^*_{\n}$.

By Lemma \ref{cont} and Lemma \ref{5.2} it follows that the topologies $\rho^*_{\LDH}$ 
and $\lambda^*_{\LDH}$ both 
coincide with $\tau_{*}$.

The following result, in a slight different form, has been proved in \cite{ak}:


\beprop
--
 $\LDH$ is complete in $\tau_*$. If $\M\in {\cal F}^R$, then $\M$ is complete in
 $\rho^*_{\m}$. Similarly, if $\N\in {\cal F}^L$, 
then $\N$ is complete in $\lambda^*_{\n}$.
\enprop
As for the third condition of Definition \ref{def3.5}, the question as to whether
 $R\LDH$ is $\rho^*_{\m}$-dense in each
 $\M \in {\cal F}^R$ (i.e. the tightness of $\LDH[\tau_{*}])$ is still open.

\vspace{3mm}
\noindent 
\underline{\it Strong* topology, $\tau_{\rm s}*$}

\noindent
 With an obvious generalization of the case of bounded operator
algebras, the strong* topology on $\L\ad(\D,\H)$ is defined by the set of
seminorms 
$$
 A \mapsto \|Af\|+\|A\ad f\|, \quad f \in \D .
$$
This topology plays a fundamental role in the study of unbounded commutants
\cite{aitrev}. 
The map $A \mapsto A\ad$ is  continuous  by definition, but   as in the
case of $\B(\H)$, the multiplications may fail to be
continuous. Therefore $\LDH[\tau_{\rm s}*]$ is {\em not} a \tpa. 
Nevertheless, the close connection between the strong* topology
and commutants allows to get some interesting density theorem. We
will show, in fact, that the set 
$$  
\B = \{ B \in \L\ad(\D,\H) : \overline{B} \in \B(\H); B \D
\subseteq \D \}
$$
is dense in $\LDH[\tau_{\rm s}*]$. This result is a consequence of
the following stronger statement.


\beprop
--
The *-algebra \F\ generated by the the identity operator and by the
set $\F(\D)$ of all finite rank operators in \D\ is dense in 
$\LDH[\tau_{\rm s}*]$ 
\enprop

\proof 
By \cite[Proposition 9]{math}, the $\tau_{\rm s}*$-closure of \F\ is 
$\F''_{\sigma\sigma}$, the weak unbounded bicommutant of \F. For
this reason it is enough to prove that $\F'_{\sigma}$ consists only of multiples
of the identity operator. Now, let $ X \in \F'_{\sigma}$; then
$X$ commutes (weakly) with each $P_\phi$, $\phi \in \D$ where
$P_\phi \psi = (\phi,\psi)\phi $. Therefore, 
$$ 
X\phi = \frac{1}{\|\phi\|^2}XP_\phi\phi = 
\frac{1}{\|\phi\|^2} P_\phi X\phi
=\frac{(\phi,X\phi)}{\|\phi\|^2}\phi.
$$ 
Now starting from two elements $\phi_1, \phi_2 \in \D$ such that
$(\phi_1, \phi_2)=0$ and using the linearity, it is easy to
show that the coefficient $\frac{(\phi,X\phi)}{\|\phi\|^2}$ does not depend on
$\phi$.
 \enproof 

\noindent
\underline{\it Weak topology, $\tau_{\rm w}$}

\noindent It is defined by the set of seminorms
$$ A \mapsto |\langle f| Ag\rangle|, \quad f,g \in \D$$

Also in this case it is readily checked that the map $A \mapsto A\ad$ is continuous. 

\noindent If $\M \in {\cal F}^R$, then the corresponding topology $\rho^{\rm w}_{\m}$ is defined
by the set of seminorms
$$
 A \in \M \to |\langle f|(X\mult A) g\rangle|, \quad X \in L\M, \, f,g \in \D.
$$
It is very easy to prove the following


\belem 
--
If $\LDH$ is self-adjoint, then the maps $A \to X\mult A$ and $A \to A\mult Y$ are 
$\tau_{\rm w}$-continuous for all 
$X \in L \LDH$ and $Y \in R \LDH$. 
\label{5.3}
\enlem
The previous lemma shows that 
$\LDH[\tau_{\rm w}]$ is a
\tpa,  at least when $\LDH$ is self-adjoint. We will now consider 
the density condition of Definition \ref{def3.5}.  In order to get results in this
direction it is useful to have at hand some information on the $\tau_{\rm w}$-continuous 
functionals on $\LDH$. In the very same way as in 
the case of weakly
continuous functionals of $\B(\H)$ (see e.g. \cite[Ch. I]{strat}), we can prove the following:


\beprop
-- 
For each $\tau_{\rm w}$-continuous linear functional $F$ on $\LDH$ 
there exist elements $f_1, \ldots, f_n;\,g_1, \ldots, g_n$ in \D\ such that
$$
 F(X) = \sum_{i=1}^n \langle f_i | Xg_i\rangle, \quad X \in \LDH .
$$
Furthermore the vectors $f_1, \ldots, f_n;\,g_1, \ldots, g_n$ can be chosen so 
that $\langle f_i|f_j\rangle =\delta_{ij}\|f_i\|^2$ and 
$\langle g_i|g_j\rangle  = \delta_{ij}\|g_i\|^2$. 
\label{form2}
\enprop
Making use of this result and of Lemma \ref{form1}, we get easily
that

\beprop
-- 
Let $\M \in {\cal F}^R$. Then, for each $\rho^{\rm w}_{\m}$-continuous linear functional $F$ on $\M$ there exist elements 
$f_1, \ldots, f_n;\,g_1, \ldots, g_n$ in \D\ and operators $A_1, \ldots A_n$ in $L\M$ such that
$$
 F(X) = \sum_{i=1}^n \langle f_i | A_i \mult X) g_i\rangle, \quad X \in \M .
$$
\enprop
We now prove the following


\beprop
-- 
$R \LDH$ is $\tau_{\rm w}$-dense in $\LDH$.
\enprop
\proof 
Were it not so, there would exist a non-zero $\tau_{\rm w}$-continuous linear functional 
$F$ on $\LDH$ which is
 zero all over $R \LDH$. By Proposition \ref{form2}, there exist elements 
$f_1, \ldots, f_n;\,g_1, \ldots, g_n$ in \D\
such that
$$
 F(X) = \sum_{i=1}^n \langle f_i | Xg_i\rangle, \quad X \in \LDH .
$$
We choose the vectors $f_1, \ldots, f_n;\,g_1, \ldots, g_n$ so that 
$\langle f_i|f_j\rangle=\delta_{ij}\|f_i\|^2$ and 
$\langle g_i|g_j\rangle=\delta_{ij}\|g_i\|^2$.

\noindent 
The finite rank operator $X$ defined by
$$
X\varphi = \sum_{j=1}^n \langle f_j | \varphi\rangle g_j, \quad \varphi \in \D
$$
clearly belongs to $R \LDH$. Then we have
$$
F(X)= \sum_{i=1}^n  \left\langle g_i \left|\sum_{i=1}^n \langle f_i
 |f_j\rangle g_j \right\rangle \right.
= \sum_{i=1}^n \|f_i\|^2\|g_i\|^2=0.
$$
This implies $f_1=\ldots f_n = g_1=\ldots g_n=0$. Therefore $F=0$ and this contradicts 
the assumption.
\enproof
\bigskip

\noindent
Unfortunately, the argument used in this proof cannot be adapted to show that 
$R \LDH$ is $\rho^{\rm w}_{\m}$-dense in each $\M \in {\cal F}^R$, so that the tightness of this \tpa\ remains to be proven.

So far we have considered only the largest \po, $\LDH$ itself. What about smaller ones? In view of the  results of Section 4, one might hope that abelian \po s would be \tpa s, possibly even tight ones. However, this is not the case, as shown by the following counterexample.

Let $T$ be a maximal symmetric operator and $\D = D(T^n), \, n < \infty$. 
Then  \cite{ait1,ait3} the \po\ generated by $T^{[1]} = T\up \D$ is the set $\gP_n(T^{[1]})$
of polynomials of degree at most $n$, powers being defined as 
$T^{[n]} = T^{[1]} \mult T^{[1]} \ldots \mult T^{[1]}$. This is an abelian, finite dimensional \po. The partial multiplication is the usual weak multiplication and $P_1 \mult P_2$ is well-defined iff 
${\rm deg}(P_1) + {\rm deg}(P_2) \leqslant n$. Thus, if $P_j$ has degree $j$, $M(P_j) = \gP_{n-j}(T^{[1]})$, so that the set of multiplier spaces is the finite scale
$$
\gP_0 \subset \gP_1 \subset \ldots \subset \gP_n, \quad 
 \gP_j \simeq \BC^{j+1}.
$$
In particular, $R\gP_n = \gP_0 = \BC$, which of course cannot be dense in
any $\gP_j$. Thus  $\gP_n(T^{[1]})$ is a (trivial) nontight \tpa.
Additional exemples of the same nature may be found in \cite{ait3}.

For a general \po\ \M, with partial multiplication $\mult$, things are easy on the algebraic side. The multipliers to be used are, of course, the internal ones, such as $R\M = R(\M) \cap \M$, and the whole lattice structure is the same as usual. However, 
the problem of topologies is  quite difficult. It is already a nontrivial problem to find the spaces of multipliers explicitly, not to speak of proving that \M\ is a \tpa! And the case of \pg s does not seem simpler.

\vspace{3mm}
\noindent {\bf Remark.} -- Once we have  endowed \D\ with the topology $t_{\L\ad}$, 
it is natural to consider 
$\L\ad(\D,\H)$ as a subspace of $\L(\D,\D')$ where $\D'$ is the 
conjugate dual of \D\ endowed with the strong dual topology $t'_{\L\ad}$.
 In this case if $A,B \in  \L\ad(\D,\H)$ then the product $A\cdot B $ always exists in 
$\L(\D,\D')$. Indeed, each $A \in  \L\ad(\D,\H)$  has an extension $\widehat{A}$ (the transposed map of $A\ad$) 
which is continuous from \H\ into $\D'$. Then $A\cdot B $ is defined
by $A\cdot B f= \widehat{A}(Bf), \; f \in \D$. The definition of the multiplication
$\cdot$ comes directly from the duality. This fact, together with Eq. (\ref{alt}) shows
that if $A\mult B$ is also well-defined, then necessarily  $A\mult
B = A\cdot B$. (This is reminiscent of the notion of weak derivative in $L^2$: given $f \in L^2$, its derivative exists always as a tempered distribution $f' \in \S'$, but $f$ belongs to the (Hilbert space) domain of $d/dx$ only if $f' \in L^2$).

One can go one step further if $\D = D^{\infty}(H)$, for some self-adjoint operator $H>1$. Then one may interpolate between $\D$ and $\D'$ by the Hilbert scale $\{ \H_n, \, n \in \BZ\}$, as discuused in Section 5.1.
The result is the same, , the partial multiplication on 
$\L(\D,\D')$  defined by continuous factorization through the spaces  $\H_n$ coincides again with the weak partial multiplication $\mult$.

\section{\hspace{-5mm}. Outcome} 

The definition of \tpa\ that emerges from this study looks quite natural, and fits well with all the examples we have given. In the abelian cases where the partial multiplication is pointwise multiplication or convolution of functions, one even gets tight \tpa s. In the more interesting case of \pa s of operators, the definition still works, but the validity of the tightness condition is generally open. It is satisfied for the `nicest'
 infinite scale, namely that built on the powers of a self-adjoint operator, but it is \emph{not} for a finite scale in general.
In fact, it is not clear how much this condition is needed. It will obviously play a role in the definition of representations by the GNS construction \cite{ait2}. When it is satisfied, it may offer interesting approximation procedures, following the standard pattern of functional analysis.

Of course, many open questions remain, in particular for \po s.
Another  challenging problem  is how to use this technique for  extending representations of \xa s to \pa s, for instance, a GNS representation. 
However, as we emphasized in the introduction, this paper is only a first step toward a general theory. Our aim was to find a structure suitable for as many significant examples as possible, and that has been obtained.
But presumably the resulting framework is too general, and one ought to specialize it to  particular cases. Clearly, more experience in this direction is needed before significant progress can be made. 

\appendix
\section*{Appendix: Proof of Proposition \ref{prop-tight_tpa}}

\section*{Acknowledgements}

This work was performed in the Institut de Physique Th\'eorique, Universit\'e Catholique de Louvain,
and the Istituto di Fisica dell' Universit\`{a} di Palermo.
We  thank both institutions for their hospitality, as well as financial support from CGRI, Communaut\'e Fran\c{c}aise de Belgique, and Ministero degli Affari Esteri, Italy.

\end{document}